\documentclass[twocolumn,traditabstract]{aa}
\usepackage{graphicx}
\usepackage{amsmath,amsfonts,amssymb}
\usepackage{txfonts}
\usepackage[breaklinks,colorlinks,citecolor=blue]{hyperref}
\usepackage{color}
\usepackage{fixltx2e}
\usepackage{natbib}
\usepackage{caption,subcaption}
\usepackage{epsfig,floatflt}
\bibpunct{(}{)}{;}{a}{}{,}

\def\setsymbol#1#2{\expandafter\def\csname #1\endcsname{#2}}
\def\getsymbol#1{\csname #1\endcsname}

\def\Planck{\textit{Planck}}

\newbox\tablebox    \newdimen\tablewidth
\def\leaderfil{\leaders\hbox to 5pt{\hss.\hss}\hfil}
\def\endPlancktablewide{\tablewidth=\textwidth 
    $$\hss\copy\tablebox\hss$$
    \vskip-\lastskip\vskip -2pt}
\def\tablenote#1 #2\par{\begingroup \parindent=0.8em
    \abovedisplayshortskip=0pt\belowdisplayshortskip=0pt
    \noindent
    $$\hss\vbox{\hsize\tablewidth \hangindent=\parindent \hangafter=1 \noindent
    \hbox to \parindent{$^#1$\hss}\strut#2\strut\par}\hss$$
    \endgroup}
\def\doubleline{\vskip 3pt\hrule \vskip 1.5pt \hrule \vskip 5pt}

\def\L2{\ifmmode L_2\else $L_2$\fi}

\def\DeltaT{\ifmmode \Delta T\else $\Delta T$\fi}
\def\deltat{\ifmmode \Delta t\else $\Delta t$\fi}
\def\fknee{\ifmmode f_{\rm knee}\else $f_{\rm knee}$\fi}
\def\Fmax{\ifmmode F_{\rm max}\else $F_{\rm max}$\fi}
\def\solar{\ifmmode{\rm M}_{\mathord\odot}\else${\rm M}_{\mathord\odot}$\fi}
\def\Msolar{\ifmmode{\rm M}_{\mathord\odot}\else${\rm M}_{\mathord\odot}$\fi}
\def\Lsolar{\ifmmode{\rm L}_{\mathord\odot}\else${\rm L}_{\mathord\odot}$\fi}

\def\inv{\ifmmode^{-1}\else$^{-1}$\fi}
\def\mo{\ifmmode^{-1}\else$^{-1}$\fi}
\def\sup#1{\ifmmode ^{\rm #1}\else $^{\rm #1}$\fi}
\def\expo#1{\ifmmode \times 10^{#1}\else $\times 10^{#1}$\fi}
\def\,{\thinspace}
\def\lsim{\mathrel{\raise .4ex\hbox{\rlap{$<$}\lower 1.2ex\hbox{$\sim$}}}}
\def\gsim{\mathrel{\raise .4ex\hbox{\rlap{$>$}\lower 1.2ex\hbox{$\sim$}}}}

\def\simprop{\mathrel{\raise .4ex\hbox{\rlap{$\propto$}\lower 1.2ex\hbox{$\sim$}}}}
\def\deg{\ifmmode^\circ\else$^\circ$\fi}
\def\pdeg{\ifmmode $\setbox0=\hbox{$^{\circ}$}\rlap{\hskip.11\wd0 .}$^{\circ}
          \else \setbox0=\hbox{$^{\circ}$}\rlap{\hskip.11\wd0 .}$^{\circ}$\fi}
\def\arcs{\ifmmode {^{\scriptstyle\prime\prime}}
          \else $^{\scriptstyle\prime\prime}$\fi}
\def\arcm{\ifmmode {^{\scriptstyle\prime}}
          \else $^{\scriptstyle\prime}$\fi}
\newdimen\sa  \newdimen\sb
\def\parcs{\sa=.07em \sb=.03em
     \ifmmode \hbox{\rlap{.}}^{\scriptstyle\prime\kern -\sb\prime}\hbox{\kern -\sa}
     \else \rlap{.}$^{\scriptstyle\prime\kern -\sb\prime}$\kern -\sa\fi}
\def\parcm{\sa=.08em \sb=.03em
     \ifmmode \hbox{\rlap{.}\kern\sa}^{\scriptstyle\prime}\hbox{\kern-\sb}
     \else \rlap{.}\kern\sa$^{\scriptstyle\prime}$\kern-\sb\fi}
\def\ra[#1 #2 #3.#4]{#1\sup{h}#2\sup{m}#3\sup{s}\llap.#4}
\def\dec[#1 #2 #3.#4]{#1\deg#2\arcm#3\arcs\llap.#4}
\def\deco[#1 #2 #3]{#1\deg#2\arcm#3\arcs}
\def\rra[#1 #2]{#1\sup{h}#2\sup{m}}

\def\dots{\relax\ifmmode \ldots\else $\ldots$\fi}
\def\WHzsr{\ifmmode $W\,Hz\mo\,sr\mo$\else W\,Hz\mo\,sr\mo\fi}
\def\mHz{\ifmmode $\,mHz$\else \,mHz\fi}
\def\GHz{\ifmmode $\,GHz$\else \,GHz\fi}
\def\mKs{\ifmmode $\,mK\,s$^{1/2}\else \,mK\,s$^{1/2}$\fi}
\def\muKs{\ifmmode \,\mu$K\,s$^{1/2}\else \,$\mu$K\,s$^{1/2}$\fi}
\def\muKRJs{\ifmmode \,\mu$K$_{\rm RJ}$\,s$^{1/2}\else \,$\mu$K$_{\rm RJ}$\,s$^{1/2}$\fi}
\def\muKHz{\ifmmode \,\mu$K\,Hz$^{-1/2}\else \,$\mu$K\,Hz$^{-1/2}$\fi}
\def\MJysr{\ifmmode \,$MJy\,sr\mo$\else \,MJy\,sr\mo\fi}
\def\MJysrmK{\ifmmode \,$MJy\,sr\mo$\,mK$_{\rm CMB}\mo\else \,MJy\,sr\mo\,mK$_{\rm CMB}\mo$\fi}
\def\microns{\ifmmode \,\mu$m$\else \,$\mu$m\fi}

\def\muK{\ifmmode \,\mu$K$\else \,$\mu$\hbox{K}\fi}
\def\microK{\ifmmode \,\mu$K$\else \,$\mu$\hbox{K}\fi}
\def\muW{\ifmmode \,\mu$W$\else \,$\mu$\hbox{W}\fi}
\def\kms{\ifmmode $\,km\,s$^{-1}\else \,km\,s$^{-1}$\fi}
\def\kmsMpc{\ifmmode $\,\kms\,Mpc\mo$\else \,\kms\,Mpc\mo\fi}
\setsymbol{LFI:center:frequency:70GHz:units}{70.3\,GHz}
\setsymbol{LFI:center:frequency:44GHz:units}{44.1\,GHz}
\setsymbol{LFI:center:frequency:30GHz:units}{28.5\,GHz}

\setsymbol{LFI:center:frequency:70GHz}{70.3}
\setsymbol{LFI:center:frequency:44GHz}{44.1}
\setsymbol{LFI:center:frequency:30GHz}{28.5}

\setsymbol{LFI:center:frequency:LFI18:Rad:M:units}{71.7\GHz}
\setsymbol{LFI:center:frequency:LFI19:Rad:M:units}{67.5\GHz}
\setsymbol{LFI:center:frequency:LFI20:Rad:M:units}{69.2\GHz}
\setsymbol{LFI:center:frequency:LFI21:Rad:M:units}{70.4\GHz}
\setsymbol{LFI:center:frequency:LFI22:Rad:M:units}{71.5\GHz}
\setsymbol{LFI:center:frequency:LFI23:Rad:M:units}{70.8\GHz}
\setsymbol{LFI:center:frequency:LFI24:Rad:M:units}{44.4\GHz}
\setsymbol{LFI:center:frequency:LFI25:Rad:M:units}{44.0\GHz}
\setsymbol{LFI:center:frequency:LFI26:Rad:M:units}{43.9\GHz}
\setsymbol{LFI:center:frequency:LFI27:Rad:M:units}{28.3\GHz}
\setsymbol{LFI:center:frequency:LFI28:Rad:M:units}{28.8\GHz}
\setsymbol{LFI:center:frequency:LFI18:Rad:S:units}{70.1\GHz}
\setsymbol{LFI:center:frequency:LFI19:Rad:S:units}{69.6\GHz}
\setsymbol{LFI:center:frequency:LFI20:Rad:S:units}{69.5\GHz}
\setsymbol{LFI:center:frequency:LFI21:Rad:S:units}{69.5\GHz}
\setsymbol{LFI:center:frequency:LFI22:Rad:S:units}{72.8\GHz}
\setsymbol{LFI:center:frequency:LFI23:Rad:S:units}{71.3\GHz}
\setsymbol{LFI:center:frequency:LFI24:Rad:S:units}{44.1\GHz}
\setsymbol{LFI:center:frequency:LFI25:Rad:S:units}{44.1\GHz}
\setsymbol{LFI:center:frequency:LFI26:Rad:S:units}{44.1\GHz}
\setsymbol{LFI:center:frequency:LFI27:Rad:S:units}{28.5\GHz}
\setsymbol{LFI:center:frequency:LFI28:Rad:S:units}{28.2\GHz}

\setsymbol{LFI:center:frequency:LFI18:Rad:M}{71.7}
\setsymbol{LFI:center:frequency:LFI19:Rad:M}{67.5}
\setsymbol{LFI:center:frequency:LFI20:Rad:M}{69.2}
\setsymbol{LFI:center:frequency:LFI21:Rad:M}{70.4}
\setsymbol{LFI:center:frequency:LFI22:Rad:M}{71.5}
\setsymbol{LFI:center:frequency:LFI23:Rad:M}{70.8}
\setsymbol{LFI:center:frequency:LFI24:Rad:M}{44.4}
\setsymbol{LFI:center:frequency:LFI25:Rad:M}{44.0}
\setsymbol{LFI:center:frequency:LFI26:Rad:M}{43.9}
\setsymbol{LFI:center:frequency:LFI27:Rad:M}{28.3}
\setsymbol{LFI:center:frequency:LFI28:Rad:M}{28.8}
\setsymbol{LFI:center:frequency:LFI18:Rad:S}{70.1}
\setsymbol{LFI:center:frequency:LFI19:Rad:S}{69.6}
\setsymbol{LFI:center:frequency:LFI20:Rad:S}{69.5}
\setsymbol{LFI:center:frequency:LFI21:Rad:S}{69.5}
\setsymbol{LFI:center:frequency:LFI22:Rad:S}{72.8}
\setsymbol{LFI:center:frequency:LFI23:Rad:S}{71.3}
\setsymbol{LFI:center:frequency:LFI24:Rad:S}{44.1}
\setsymbol{LFI:center:frequency:LFI25:Rad:S}{44.1}
\setsymbol{LFI:center:frequency:LFI26:Rad:S}{44.1}
\setsymbol{LFI:center:frequency:LFI27:Rad:S}{28.5}
\setsymbol{LFI:center:frequency:LFI28:Rad:S}{28.2}

\setsymbol{LFI:white:noise:sensitivity:70GHz:units}{134.7\muKs}
\setsymbol{LFI:white:noise:sensitivity:44GHz:units}{164.7\muKs}
\setsymbol{LFI:white:noise:sensitivity:30GHz:units}{143.4\muKs}

\setsymbol{LFI:white:noise:sensitivity:70GHz}{134.7}
\setsymbol{LFI:white:noise:sensitivity:44GHz}{164.7}
\setsymbol{LFI:white:noise:sensitivity:30GHz}{143.4}

\setsymbol{LFI:white:noise:sensitivity:LFI18:Rad:M:units}{512.0\muKs}
\setsymbol{LFI:white:noise:sensitivity:LFI19:Rad:M:units}{581.4\muKs}
\setsymbol{LFI:white:noise:sensitivity:LFI20:Rad:M:units}{590.8\muKs}
\setsymbol{LFI:white:noise:sensitivity:LFI21:Rad:M:units}{455.2\muKs}
\setsymbol{LFI:white:noise:sensitivity:LFI22:Rad:M:units}{492.0\muKs}
\setsymbol{LFI:white:noise:sensitivity:LFI23:Rad:M:units}{507.7\muKs}
\setsymbol{LFI:white:noise:sensitivity:LFI24:Rad:M:units}{462.2\muKs}
\setsymbol{LFI:white:noise:sensitivity:LFI25:Rad:M:units}{413.6\muKs}
\setsymbol{LFI:white:noise:sensitivity:LFI26:Rad:M:units}{478.6\muKs}
\setsymbol{LFI:white:noise:sensitivity:LFI27:Rad:M:units}{277.7\muKs}
\setsymbol{LFI:white:noise:sensitivity:LFI28:Rad:M:units}{312.3\muKs}
\setsymbol{LFI:white:noise:sensitivity:LFI18:Rad:S:units}{465.7\muKs}
\setsymbol{LFI:white:noise:sensitivity:LFI19:Rad:S:units}{555.6\muKs}
\setsymbol{LFI:white:noise:sensitivity:LFI20:Rad:S:units}{623.2\muKs}
\setsymbol{LFI:white:noise:sensitivity:LFI21:Rad:S:units}{564.1\muKs}
\setsymbol{LFI:white:noise:sensitivity:LFI22:Rad:S:units}{534.4\muKs}
\setsymbol{LFI:white:noise:sensitivity:LFI23:Rad:S:units}{542.4\muKs}
\setsymbol{LFI:white:noise:sensitivity:LFI24:Rad:S:units}{399.2\muKs}
\setsymbol{LFI:white:noise:sensitivity:LFI25:Rad:S:units}{392.6\muKs}
\setsymbol{LFI:white:noise:sensitivity:LFI26:Rad:S:units}{418.6\muKs}
\setsymbol{LFI:white:noise:sensitivity:LFI27:Rad:S:units}{302.9\muKs}
\setsymbol{LFI:white:noise:sensitivity:LFI28:Rad:S:units}{285.3\muKs}

\setsymbol{LFI:white:noise:sensitivity:LFI18:Rad:M}{512.0}
\setsymbol{LFI:white:noise:sensitivity:LFI19:Rad:M}{581.4}
\setsymbol{LFI:white:noise:sensitivity:LFI20:Rad:M}{590.8}
\setsymbol{LFI:white:noise:sensitivity:LFI21:Rad:M}{455.2}
\setsymbol{LFI:white:noise:sensitivity:LFI22:Rad:M}{492.0}
\setsymbol{LFI:white:noise:sensitivity:LFI23:Rad:M}{507.7}
\setsymbol{LFI:white:noise:sensitivity:LFI24:Rad:M}{462.2}
\setsymbol{LFI:white:noise:sensitivity:LFI25:Rad:M}{413.6}
\setsymbol{LFI:white:noise:sensitivity:LFI26:Rad:M}{478.6}
\setsymbol{LFI:white:noise:sensitivity:LFI27:Rad:M}{277.7}
\setsymbol{LFI:white:noise:sensitivity:LFI28:Rad:M}{312.3}
\setsymbol{LFI:white:noise:sensitivity:LFI18:Rad:S}{465.7}
\setsymbol{LFI:white:noise:sensitivity:LFI19:Rad:S}{555.6}
\setsymbol{LFI:white:noise:sensitivity:LFI20:Rad:S}{623.2}
\setsymbol{LFI:white:noise:sensitivity:LFI21:Rad:S}{564.1}
\setsymbol{LFI:white:noise:sensitivity:LFI22:Rad:S}{534.4}
\setsymbol{LFI:white:noise:sensitivity:LFI23:Rad:S}{542.4}
\setsymbol{LFI:white:noise:sensitivity:LFI24:Rad:S}{399.2}
\setsymbol{LFI:white:noise:sensitivity:LFI25:Rad:S}{392.6}
\setsymbol{LFI:white:noise:sensitivity:LFI26:Rad:S}{418.6}
\setsymbol{LFI:white:noise:sensitivity:LFI27:Rad:S}{302.9}
\setsymbol{LFI:white:noise:sensitivity:LFI28:Rad:S}{285.3}

\setsymbol{LFI:knee:frequency:70GHz:units}{29.5\mHz}
\setsymbol{LFI:knee:frequency:44GHz:units}{56.2\mHz}
\setsymbol{LFI:knee:frequency:30GHz:units}{113.7\mHz}

\setsymbol{LFI:knee:frequency:70GHz}{29.5}
\setsymbol{LFI:knee:frequency:44GHz}{56.2}
\setsymbol{LFI:knee:frequency:30GHz}{113.7}

\setsymbol{LFI:knee:frequency:LFI18:Rad:M:units}{16.3\mHz}
\setsymbol{LFI:knee:frequency:LFI19:Rad:M:units}{15.1\mHz}
\setsymbol{LFI:knee:frequency:LFI20:Rad:M:units}{18.7\mHz}
\setsymbol{LFI:knee:frequency:LFI21:Rad:M:units}{37.2\mHz}
\setsymbol{LFI:knee:frequency:LFI22:Rad:M:units}{12.7\mHz}
\setsymbol{LFI:knee:frequency:LFI23:Rad:M:units}{34.6\mHz}
\setsymbol{LFI:knee:frequency:LFI24:Rad:M:units}{46.2\mHz}
\setsymbol{LFI:knee:frequency:LFI25:Rad:M:units}{24.9\mHz}
\setsymbol{LFI:knee:frequency:LFI26:Rad:M:units}{67.6\mHz}
\setsymbol{LFI:knee:frequency:LFI27:Rad:M:units}{187.4\mHz}
\setsymbol{LFI:knee:frequency:LFI28:Rad:M:units}{122.2\mHz}
\setsymbol{LFI:knee:frequency:LFI18:Rad:S:units}{17.7\mHz}
\setsymbol{LFI:knee:frequency:LFI19:Rad:S:units}{22.0\mHz}
\setsymbol{LFI:knee:frequency:LFI20:Rad:S:units}{8.7\mHz}
\setsymbol{LFI:knee:frequency:LFI21:Rad:S:units}{25.9\mHz}
\setsymbol{LFI:knee:frequency:LFI22:Rad:S:units}{15.8\mHz}
\setsymbol{LFI:knee:frequency:LFI23:Rad:S:units}{129.8\mHz}
\setsymbol{LFI:knee:frequency:LFI24:Rad:S:units}{100.9\mHz}
\setsymbol{LFI:knee:frequency:LFI25:Rad:S:units}{38.9\mHz}
\setsymbol{LFI:knee:frequency:LFI26:Rad:S:units}{58.9\mHz}
\setsymbol{LFI:knee:frequency:LFI27:Rad:S:units}{104.4\mHz}
\setsymbol{LFI:knee:frequency:LFI28:Rad:S:units}{40.7\mHz}

\setsymbol{LFI:knee:frequency:LFI18:Rad:M}{16.3}
\setsymbol{LFI:knee:frequency:LFI19:Rad:M}{15.1}
\setsymbol{LFI:knee:frequency:LFI20:Rad:M}{18.7}
\setsymbol{LFI:knee:frequency:LFI21:Rad:M}{37.2}
\setsymbol{LFI:knee:frequency:LFI22:Rad:M}{12.7}
\setsymbol{LFI:knee:frequency:LFI23:Rad:M}{34.6}
\setsymbol{LFI:knee:frequency:LFI24:Rad:M}{46.2}
\setsymbol{LFI:knee:frequency:LFI25:Rad:M}{24.9}
\setsymbol{LFI:knee:frequency:LFI26:Rad:M}{67.6}
\setsymbol{LFI:knee:frequency:LFI27:Rad:M}{187.4}
\setsymbol{LFI:knee:frequency:LFI28:Rad:M}{122.2}
\setsymbol{LFI:knee:frequency:LFI18:Rad:S}{17.7}
\setsymbol{LFI:knee:frequency:LFI19:Rad:S}{22.0}
\setsymbol{LFI:knee:frequency:LFI20:Rad:S}{8.7}
\setsymbol{LFI:knee:frequency:LFI21:Rad:S}{25.9}
\setsymbol{LFI:knee:frequency:LFI22:Rad:S}{15.8}
\setsymbol{LFI:knee:frequency:LFI23:Rad:S}{129.8}
\setsymbol{LFI:knee:frequency:LFI24:Rad:S}{100.9}
\setsymbol{LFI:knee:frequency:LFI25:Rad:S}{38.9}
\setsymbol{LFI:knee:frequency:LFI26:Rad:S}{58.9}
\setsymbol{LFI:knee:frequency:LFI27:Rad:S}{104.4}
\setsymbol{LFI:knee:frequency:LFI28:Rad:S}{40.7}

\setsymbol{LFI:slope:70GHz:units}{$-1.03$\mHz}
\setsymbol{LFI:slope:44GHz:units}{$-0.89$\mHz}
\setsymbol{LFI:slope:30GHz:units}{$-0.87$\mHz}

\setsymbol{LFI:slope:70GHz}{$-1.03$}
\setsymbol{LFI:slope:44GHz}{$-0.89$}
\setsymbol{LFI:slope:30GHz}{$-0.87$}

\setsymbol{LFI:slope:LFI18:Rad:M:units}{$-1.04$\mHz}
\setsymbol{LFI:slope:LFI19:Rad:M:units}{$-1.09$\mHz}
\setsymbol{LFI:slope:LFI20:Rad:M:units}{$-0.69$\mHz}
\setsymbol{LFI:slope:LFI21:Rad:M:units}{$-1.56$\mHz}
\setsymbol{LFI:slope:LFI22:Rad:M:units}{$-1.01$\mHz}
\setsymbol{LFI:slope:LFI23:Rad:M:units}{$-0.96$\mHz}
\setsymbol{LFI:slope:LFI24:Rad:M:units}{$-0.83$\mHz}
\setsymbol{LFI:slope:LFI25:Rad:M:units}{$-0.91$\mHz}
\setsymbol{LFI:slope:LFI26:Rad:M:units}{$-0.95$\mHz}
\setsymbol{LFI:slope:LFI27:Rad:M:units}{$-0.87$\mHz}
\setsymbol{LFI:slope:LFI28:Rad:M:units}{$-0.88$\mHz}
\setsymbol{LFI:slope:LFI18:Rad:S:units}{$-1.15$\mHz}
\setsymbol{LFI:slope:LFI19:Rad:S:units}{$-1.00$\mHz}
\setsymbol{LFI:slope:LFI20:Rad:S:units}{$-0.95$\mHz}
\setsymbol{LFI:slope:LFI21:Rad:S:units}{$-0.92$\mHz}
\setsymbol{LFI:slope:LFI22:Rad:S:units}{$-1.01$\mHz}
\setsymbol{LFI:slope:LFI23:Rad:S:units}{$-0.95$\mHz}
\setsymbol{LFI:slope:LFI24:Rad:S:units}{$-0.73$\mHz}
\setsymbol{LFI:slope:LFI25:Rad:S:units}{$-1.16$\mHz}
\setsymbol{LFI:slope:LFI26:Rad:S:units}{$-0.79$\mHz}
\setsymbol{LFI:slope:LFI27:Rad:S:units}{$-0.82$\mHz}
\setsymbol{LFI:slope:LFI28:Rad:S:units}{$-0.91$\mHz}

\setsymbol{LFI:slope:LFI18:Rad:M}{$-1.04$}
\setsymbol{LFI:slope:LFI19:Rad:M}{$-1.09$}
\setsymbol{LFI:slope:LFI20:Rad:M}{$-0.69$}
\setsymbol{LFI:slope:LFI21:Rad:M}{$-1.56$}
\setsymbol{LFI:slope:LFI22:Rad:M}{$-1.01$}
\setsymbol{LFI:slope:LFI23:Rad:M}{$-0.96$}
\setsymbol{LFI:slope:LFI24:Rad:M}{$-0.83$}
\setsymbol{LFI:slope:LFI25:Rad:M}{$-0.91$}
\setsymbol{LFI:slope:LFI26:Rad:M}{$-0.95$}
\setsymbol{LFI:slope:LFI27:Rad:M}{$-0.87$}
\setsymbol{LFI:slope:LFI28:Rad:M}{$-0.88$}
\setsymbol{LFI:slope:LFI18:Rad:S}{$-1.15$}
\setsymbol{LFI:slope:LFI19:Rad:S}{$-1.00$}
\setsymbol{LFI:slope:LFI20:Rad:S}{$-0.95$}
\setsymbol{LFI:slope:LFI21:Rad:S}{$-0.92$}
\setsymbol{LFI:slope:LFI22:Rad:S}{$-1.01$}
\setsymbol{LFI:slope:LFI23:Rad:S}{$-0.95$}
\setsymbol{LFI:slope:LFI24:Rad:S}{$-0.73$}
\setsymbol{LFI:slope:LFI25:Rad:S}{$-1.16$}
\setsymbol{LFI:slope:LFI26:Rad:S}{$-0.79$}
\setsymbol{LFI:slope:LFI27:Rad:S}{$-0.82$}
\setsymbol{LFI:slope:LFI28:Rad:S}{$-0.91$}

\setsymbol{LFI:FWHM:70GHz:units}{13\parcm01}
\setsymbol{LFI:FWHM:44GHz:units}{27\parcm92}
\setsymbol{LFI:FWHM:30GHz:units}{32\parcm65}

\setsymbol{LFI:FWHM:70GHz}{13.01}
\setsymbol{LFI:FWHM:44GHz}{27.92}
\setsymbol{LFI:FWHM:30GHz}{32.65}

\setsymbol{LFI:FWHM:LFI18:units}{13\parcm39}
\setsymbol{LFI:FWHM:LFI19:units}{13\parcm01}
\setsymbol{LFI:FWHM:LFI20:units}{12\parcm75}
\setsymbol{LFI:FWHM:LFI21:units}{12\parcm74}
\setsymbol{LFI:FWHM:LFI22:units}{12\parcm87}
\setsymbol{LFI:FWHM:LFI23:units}{13\parcm27}
\setsymbol{LFI:FWHM:LFI24:units}{22\parcm98}
\setsymbol{LFI:FWHM:LFI25:units}{30\parcm46}
\setsymbol{LFI:FWHM:LFI26:units}{30\parcm31}
\setsymbol{LFI:FWHM:LFI27:units}{32\parcm65}
\setsymbol{LFI:FWHM:LFI28:units}{32\parcm66}

\setsymbol{LFI:FWHM:LFI18}{13.39}
\setsymbol{LFI:FWHM:LFI19}{13.01}
\setsymbol{LFI:FWHM:LFI20}{12.75}
\setsymbol{LFI:FWHM:LFI21}{12.74}
\setsymbol{LFI:FWHM:LFI22}{12.87}
\setsymbol{LFI:FWHM:LFI23}{13.27}
\setsymbol{LFI:FWHM:LFI24}{22.98}
\setsymbol{LFI:FWHM:LFI25}{30.46}
\setsymbol{LFI:FWHM:LFI26}{30.31}
\setsymbol{LFI:FWHM:LFI27}{32.65}
\setsymbol{LFI:FWHM:LFI28}{32.66}

\setsymbol{LFI:FWHM:uncertainty:LFI18:units}{0.170\arcm}
\setsymbol{LFI:FWHM:uncertainty:LFI19:units}{0.174\arcm}
\setsymbol{LFI:FWHM:uncertainty:LFI20:units}{0.170\arcm}
\setsymbol{LFI:FWHM:uncertainty:LFI21:units}{0.156\arcm}
\setsymbol{LFI:FWHM:uncertainty:LFI22:units}{0.164\arcm}
\setsymbol{LFI:FWHM:uncertainty:LFI23:units}{0.171\arcm}
\setsymbol{LFI:FWHM:uncertainty:LFI24:units}{0.652\arcm}
\setsymbol{LFI:FWHM:uncertainty:LFI25:units}{1.075\arcm}
\setsymbol{LFI:FWHM:uncertainty:LFI26:units}{1.131\arcm}
\setsymbol{LFI:FWHM:uncertainty:LFI27:units}{1.266\arcm}
\setsymbol{LFI:FWHM:uncertainty:LFI28:units}{1.287\arcm}

\setsymbol{LFI:FWHM:uncertainty:LFI18}{0.170}
\setsymbol{LFI:FWHM:uncertainty:LFI19}{0.174}
\setsymbol{LFI:FWHM:uncertainty:LFI20}{0.170}
\setsymbol{LFI:FWHM:uncertainty:LFI21}{0.156}
\setsymbol{LFI:FWHM:uncertainty:LFI22}{0.164}
\setsymbol{LFI:FWHM:uncertainty:LFI23}{0.171}
\setsymbol{LFI:FWHM:uncertainty:LFI24}{0.652}
\setsymbol{LFI:FWHM:uncertainty:LFI25}{1.075}
\setsymbol{LFI:FWHM:uncertainty:LFI26}{1.131}
\setsymbol{LFI:FWHM:uncertainty:LFI27}{1.266}
\setsymbol{LFI:FWHM:uncertainty:LFI28}{1.287}

\setsymbol{HFI:center:frequency:100GHz:units}{100\,GHz}
\setsymbol{HFI:center:frequency:143GHz:units}{143\,GHz}
\setsymbol{HFI:center:frequency:217GHz:units}{217\,GHz}
\setsymbol{HFI:center:frequency:353GHz:units}{353\,GHz}
\setsymbol{HFI:center:frequency:545GHz:units}{545\,GHz}
\setsymbol{HFI:center:frequency:857GHz:units}{857\,GHz}

\setsymbol{HFI:center:frequency:100GHz}{100}
\setsymbol{HFI:center:frequency:143GHz}{143}
\setsymbol{HFI:center:frequency:217GHz}{217}
\setsymbol{HFI:center:frequency:353GHz}{353}
\setsymbol{HFI:center:frequency:545GHz}{545}
\setsymbol{HFI:center:frequency:857GHz}{857}

\setsymbol{HFI:Ndetectors:100GHz}{8}
\setsymbol{HFI:Ndetectors:143GHz}{11}
\setsymbol{HFI:Ndetectors:217GHz}{12}
\setsymbol{HFI:Ndetectors:353GHz}{12}
\setsymbol{HFI:Ndetectors:545GHz}{3}
\setsymbol{HFI:Ndetectors:857GHz}{4}

\setsymbol{HFI:FWHM:Maps:100GHz:units}{9\parcm88}
\setsymbol{HFI:FWHM:Maps:143GHz:units}{7\parcm18}
\setsymbol{HFI:FWHM:Maps:217GHz:units}{4\parcm87}
\setsymbol{HFI:FWHM:Maps:353GHz:units}{4\parcm65}
\setsymbol{HFI:FWHM:Maps:545GHz:units}{4\parcm72}
\setsymbol{HFI:FWHM:Maps:857GHz:units}{4\parcm39}
\setsymbol{HFI:FWHM:Maps:100GHz}{9.88}
\setsymbol{HFI:FWHM:Maps:143GHz}{7.18}
\setsymbol{HFI:FWHM:Maps:217GHz}{4.87}
\setsymbol{HFI:FWHM:Maps:353GHz}{4.65}
\setsymbol{HFI:FWHM:Maps:545GHz}{4.72}
\setsymbol{HFI:FWHM:Maps:857GHz}{4.39}

\setsymbol{HFI:beam:ellipticity:Maps:100GHz}{1.15}
\setsymbol{HFI:beam:ellipticity:Maps:143GHz}{1.01}
\setsymbol{HFI:beam:ellipticity:Maps:217GHz}{1.06}
\setsymbol{HFI:beam:ellipticity:Maps:353GHz}{1.05}
\setsymbol{HFI:beam:ellipticity:Maps:545GHz}{1.14}
\setsymbol{HFI:beam:ellipticity:Maps:857GHz}{1.19}

\setsymbol{HFI:FWHM:Mars:100GHz:units}{9\parcm37}
\setsymbol{HFI:FWHM:Mars:143GHz:units}{7\parcm04}
\setsymbol{HFI:FWHM:Mars:217GHz:units}{4\parcm68}
\setsymbol{HFI:FWHM:Mars:353GHz:units}{4\parcm43}
\setsymbol{HFI:FWHM:Mars:545GHz:units}{3\parcm80}
\setsymbol{HFI:FWHM:Mars:857GHz:units}{3\parcm67}

\setsymbol{HFI:FWHM:Mars:100GHz}{9.37}
\setsymbol{HFI:FWHM:Mars:143GHz}{7.04}
\setsymbol{HFI:FWHM:Mars:217GHz}{4.68}
\setsymbol{HFI:FWHM:Mars:353GHz}{4.43}
\setsymbol{HFI:FWHM:Mars:545GHz}{3.80}
\setsymbol{HFI:FWHM:Mars:857GHz}{3.67}

\setsymbol{HFI:beam:ellipticity:Mars:100GHz}{1.18}
\setsymbol{HFI:beam:ellipticity:Mars:143GHz}{1.03}
\setsymbol{HFI:beam:ellipticity:Mars:217GHz}{1.14}
\setsymbol{HFI:beam:ellipticity:Mars:353GHz}{1.09}
\setsymbol{HFI:beam:ellipticity:Mars:545GHz}{1.25}
\setsymbol{HFI:beam:ellipticity:Mars:857GHz}{1.03}

\setsymbol{HFI:CMB:relative:calibration:100GHz}{$\lsim 1\%$}
\setsymbol{HFI:CMB:relative:calibration:143GHz}{$\lsim 1\%$}
\setsymbol{HFI:CMB:relative:calibration:217GHz}{$\lsim 1\%$}
\setsymbol{HFI:CMB:relative:calibration:353GHz}{$\lsim 1\%$}
\setsymbol{HFI:CMB:relative:calibration:545GHz}{}
\setsymbol{HFI:CMB:relative:calibration:857GHz}{}

\setsymbol{HFI:CMB:absolute:calibration:100GHz}{$\lsim 2\%$}
\setsymbol{HFI:CMB:absolute:calibration:143GHz}{$\lsim 2\%$}
\setsymbol{HFI:CMB:absolute:calibration:217GHz}{$\lsim 2\%$}
\setsymbol{HFI:CMB:absolute:calibration:353GHz}{$\lsim 2\%$}
\setsymbol{HFI:CMB:absolute:calibration:545GHz}{}
\setsymbol{HFI:CMB:absolute:calibration:857GHz}{}

\setsymbol{HFI:FIRAS:gain:calibration:accuracy:statistical:100GHz}{}
\setsymbol{HFI:FIRAS:gain:calibration:accuracy:statistical:143GHz}{}
\setsymbol{HFI:FIRAS:gain:calibration:accuracy:statistical:217GHz}{}
\setsymbol{HFI:FIRAS:gain:calibration:accuracy:statistical:353GHz}{2.5\%}
\setsymbol{HFI:FIRAS:gain:calibration:accuracy:statistical:545GHz}{1\%}
\setsymbol{HFI:FIRAS:gain:calibration:accuracy:statistical:857GHz}{0.5\%}

\setsymbol{HFI:FIRAS:gain:calibration:accuracy:systematic:100GHz}{}
\setsymbol{HFI:FIRAS:gain:calibration:accuracy:systematic:143GHz}{}
\setsymbol{HFI:FIRAS:gain:calibration:accuracy:systematic:217GHz}{}
\setsymbol{HFI:FIRAS:gain:calibration:accuracy:systematic:353GHz}{}
\setsymbol{HFI:FIRAS:gain:calibration:accuracy:systematic:545GHz}{7\%}
\setsymbol{HFI:FIRAS:gain:calibration:accuracy:systematic:857GHz}{7\%}

\setsymbol{HFI:FIRAS:zero:point:accuracy:100GHz:units}{0.8\MJysr}
\setsymbol{HFI:FIRAS:zero:point:accuracy:143GHz:units}{}
\setsymbol{HFI:FIRAS:zero:point:accuracy:217GHz:units}{}
\setsymbol{HFI:FIRAS:zero:point:accuracy:353GHz:units}{1.4\MJysr}
\setsymbol{HFI:FIRAS:zero:point:accuracy:545GHz:units}{2.2\MJysr}
\setsymbol{HFI:FIRAS:zero:point:accuracy:857GHz:units}{1.7\MJysr}

\setsymbol{HFI:FIRAS:zero:point:accuracy:100GHz}{0.8}
\setsymbol{HFI:FIRAS:zero:point:accuracy:143GHz}{}
\setsymbol{HFI:FIRAS:zero:point:accuracy:217GHz}{}
\setsymbol{HFI:FIRAS:zero:point:accuracy:353GHz}{1.4}
\setsymbol{HFI:FIRAS:zero:point:accuracy:545GHz}{2.2}
\setsymbol{HFI:FIRAS:zero:point:accuracy:857GHz}{1.7}

\setsymbol{HFI:unit:conversion:100GHz:units}{0.2415\MJysrmK}
\setsymbol{HFI:unit:conversion:143GHz:units}{0.3694\MJysrmK}
\setsymbol{HFI:unit:conversion:217GHz:units}{0.4811\MJysrmK}
\setsymbol{HFI:unit:conversion:353GHz:units}{0.2883\MJysrmK}
\setsymbol{HFI:unit:conversion:545GHz:units}{0.05826\MJysrmK}
\setsymbol{HFI:unit:conversion:857GHz:units}{0.002238\MJysrmK}

\setsymbol{HFI:unit:conversion:100GHz}{0.2415}
\setsymbol{HFI:unit:conversion:143GHz}{0.3694}
\setsymbol{HFI:unit:conversion:217GHz}{0.4811}
\setsymbol{HFI:unit:conversion:353GHz}{0.2883}
\setsymbol{HFI:unit:conversion:545GHz}{0.05826}
\setsymbol{HFI:unit:conversion:857GHz}{0.002238}

\setsymbol{HFI:colour:correction:alpha=-2:V1.01:100GHz}{0.9893}
\setsymbol{HFI:colour:correction:alpha=-2:V1.01:143GHz}{0.9759}
\setsymbol{HFI:colour:correction:alpha=-2:V1.01:217GHz}{1.0007}
\setsymbol{HFI:colour:correction:alpha=-2:V1.01:353GHz}{1.0028}
\setsymbol{HFI:colour:correction:alpha=-2:V1.01:545GHz}{1.0019}
\setsymbol{HFI:colour:correction:alpha=-2:V1.01:857GHz}{0.9889}

\setsymbol{HFI:colour:correction:alpha=0:V1.01:100GHz}{1.0008}
\setsymbol{HFI:colour:correction:alpha=0:V1.01:143GHz}{1.0148}
\setsymbol{HFI:colour:correction:alpha=0:V1.01:217GHz}{0.9909}
\setsymbol{HFI:colour:correction:alpha=0:V1.01:353GHz}{0.9888}
\setsymbol{HFI:colour:correction:alpha=0:V1.01:545GHz}{0.9878}
\setsymbol{HFI:colour:correction:alpha=0:V1.01:857GHz}{1.0014}

\def\COBE{\textit{COBE}}

\def\healpix{\texttt{HEALPix}}
\def\commander{\texttt{Commander}}

\renewcommand{\a}[0]{\mathbf{a}}
\newcommand{\m}[0]{\mathbf{m}}

\renewcommand{\L}[0]{\mathbf{L}}

\def\inv{^{-1}}

\title{Monopole and dipole estimation for multi-frequency sky maps by
  linear regression}

\author{\small
I.~K.~Wehus\inst{1}
\and
U.~Fuskeland\inst{2}
\and
H.~K.~Eriksen\inst{2}
\and
A.~J.~Banday\inst{3, 4}
\and
C.~Dickinson\inst{5}
\and
T.~Ghosh\inst{6}
\and
K.~M.~G\'{o}rski\inst{1, 7}
\and
C.~R.~Lawrence\inst{1}
\and
J.~P.~Leahy\inst{5}
\and
D.~Maino\inst{8, 9}
\and
P.~Reich\inst{10}
\and
W.~Reich\inst{10}
}
\institute{\small
1. Jet Propulsion Laboratory, California Institute of Technology, 4800 Oak Grove Drive, Pasadena, California, U.S.A., \thanks{\email{Ingunn.K.Wehus@jpl.nasa.gov}}\\
2. Institute of Theoretical Astrophysics, University of Oslo, Blindern, Oslo, Norway\\
3. Universit\'{e} de Toulouse, UPS-OMP, IRAP, F-31028 Toulouse cedex 4, France\\
4. CNRS, IRAP, 9 Av. colonel Roche, BP 44346, F-31028 Toulouse cedex 4, France\\
5. Jodrell Bank Centre for Astrophysics, Alan Turing Building, School of Physics and Astronomy, \\
\hspace*{0.3cm}The University of Manchester, Oxford Road, Manchester, M13 9PL, U.K.\\
6. Institut d'Astrophysique Spatiale, CNRS (UMR8617) Universit\'{e} Paris-Sud 11, B\^{a}timent 121, Orsay, France\\
7. Warsaw University Observatory, Aleje Ujazdowskie 4, 00-478 Warszawa, Poland\\
8. Dipartimento di Fisica, Universit\`{a} degli Studi di Milano, Via Celoria, 16, Milano, Italy\\
9. INAF/IASF Milano, Via E. Bassini 15, Milano, Italy\\
10. Max-Planck-Institut f\"{u}r Radioastronomie, Auf dem H\"{u}gel 69, 53121 Bonn, Germany \\
}

\begin{document}

\abstract{ We describe a simple but efficient method for deriving a
  consistent set of monopole and dipole corrections for
  multi-frequency sky map data sets, allowing robust parametric
  component separation with the same data set. The computational core
  of this method is linear regression between pairs of frequency maps,
  often called ``T-T plots''. Individual contributions from monopole
  and dipole terms are determined by performing the regression locally
  in patches on the sky, while the degeneracy between different
  frequencies is lifted whenever the dominant foreground component
  exhibits a significant spatial spectral index variation. Based on
  this method, we present two different, but each internally
  consistent, sets of monopole and dipole coefficients for the 9-year
  WMAP, \Planck\ 2013, SFD $100\,\mu\textrm{m}$, Haslam 408$\,$MHz and
  Reich \& Reich 1420$\,$MHz maps. The two sets have been derived with
  different analysis assumptions and data selection, and provides an
  estimate of residual systematic uncertainties. In general, our values are
  in good agreement with previously published results. Among the most
  notable results are a relative dipole between the WMAP and
  \Planck\ experiments of 10--15$\,\mu\textrm{K}$ (depending on
  frequency), an estimate of the 408$\,$MHz map monopole of
  $8.9\pm1.3\,\textrm{K}$, and a non-zero dipole in the 1420$\,$MHz map
  of $0.15\pm0.03\,\textrm{K}$ pointing towards Galactic coordinates
  $(l,b) = (308^{\circ},-36^{\circ})\pm14^{\circ}$. These values
  represent the sum of any instrumental and data processing offsets,
  as well as any Galactic or extra-Galactic component that is
  spectrally uniform over the full sky. }


\maketitle

\section{Introduction}
\label{sec:introduction}

The cosmic microwave background (CMB) fluctuations consist of small
variations with a root-mean-square (RMS) of $70\,\mu\textrm{K}$
imprinted on top of a mean temperature of 2.73$\,$K and a
Doppler-induced dipole of $\sim$$3\,\textrm{mK}$. These minute
variations thus correspond to fractional fluctuations at the level of
$\sim$$10^{-5}$ relative to the total signal. To minimize systematic
uncertainties, modern CMB anisotropy experiments are therefore forced
to employ some form of differential measuring technique, eliminating
the large 2.73$\,$K offset already at the instrument level. Both
\COBE-DMR \citep{smoot:1992} and WMAP \citep{bennett:2013} employed
coupled differencing assemblies that only recorded temperature
differences between two positions on the sky, while for \Planck\ the
instrumental offset is large and unknown, and cannot be used to
constrain the monopole \citep{planck_mission}. However, while
necessary for systematics suppression, this also implies that these
experiments are intrinsically unable to measure the true absolute
zero-point (or monopole) of their final maps. In addition, the dipole
is also associated with a large relative uncertainty because of the
large numerical value of the CMB Doppler dipole; a small relative
error in the determination of the CMB dipole direction can induce a
dipole error of many microkelvins in a CMB map. Typically, this will
be strongly correlated among frequencies within a single experiment,
though, and so informative priors can be imposed among frequency
channels within a given experiment. For foreground-dominated frequency
channels, instrumental systematics, such as gain fluctuations, may
induce dipole errors.

Removing the monopole and dipole from CMB data sets does not
constitute a major limitation in terms of CMB-based cosmological
parameter analysis, since losing a handful harmonic modes out of many
thousands only negligibly reduces the total amount of available
information. However, it does have a significant indirect impact
because of the presence of non-cosmological foreground contamination
from Galactic and extra-Galactic sources. In order to obtain a clean
image of the cosmological CMB fluctuations, this foreground
contamination has to be removed from the raw sky maps through some
form of component separation prior to power spectrum and parameter
estimation. A wide range of such methods have been already been
proposed \citep[e.g.,][and references therein]{planck_compsep},
ranging from the simplest of template fitting and internal linear
combination approaches through blind- or semi-blind image processing
techniques to full-blown parametric Bayesian methods employing
physical models. Except for the very simplest methods, all of these
exploit the fact that all foreground frequency spectra are
qualitatively different from the CMB spectrum. For instance, while the
CMB spectrum is that of a perfect blackbody, thermal dust emission can
be well approximated by that of a one- or two-component
greybody. However, such relations clearly only hold if there are no
arbitrary offsets between the different frequency maps. In other
words, spurious monopole and dipole errors can bias any estimation
algorithm that exploits frequency dependencies, and this can in turn
lead to leakage between various components, and eventually
contamination in the CMB estimate.

A number of methods have already been proposed in the literature for
estimating monopoles, while fewer have addressed residual dipoles. One
example of the former is the co-secant method adopted by the WMAP team
\citep{bennett:2003}. In this case, the Galactic signal is
approximated as plane-parallel in Galactic coordinates, with an
amplitude falling roughly proportionally with the co-secant of the
latitude. The major weakness of this method is that the Galaxy is
neither plane-parallel nor follows a co-secant, and the method does
also not account for residual dipole terms. A second approach was
proposed by \citet{eriksen:2008}, who include the monopole and dipole
terms as additional free parameters within a global Bayesian
parametric framework. The major weakness with this method is a large
degeneracy between the monopole and dipole terms relative to the
unknown zero-point of each foreground; it is possible to add a
constant to each foreground amplitude, and then subtract a
corresponding frequency-scaled offset from each monopole, leaving the
net sum unchanged.

A third widely used method for setting the zero-level of radio maps is
that of linear regression, or through so-called ``T-T plots''. This
method has a long and prominent history in radio astronomy \citep[see,
  e.g.,][and references
  therein]{turtle:1962,davies:1996,reich:1988,reich:2004,wehus:2013}, as it
provides a highly robust estimate of the spectral index of a single
signal component given observations at two different frequencies. When
plotting the measured pixel values at one frequency as a function of
the measured pixel values at the other frequency, the spectral index
is given (up to a constant factor) by the \emph{slope} of the
resulting T-T plot, which is easily found by linear regression. The
main virtue of this spectral index estimate is that it is completely
insensitive to any constant offset in either of the two frequency
maps, since these only affect the regression intercept, not the
slope. Intuitively, offsets only shift the scatter plot horizontally
or vertically, but they do not deform or rotate it.

\begin{figure}[t]
\begin{center}
\mbox{\epsfig{figure=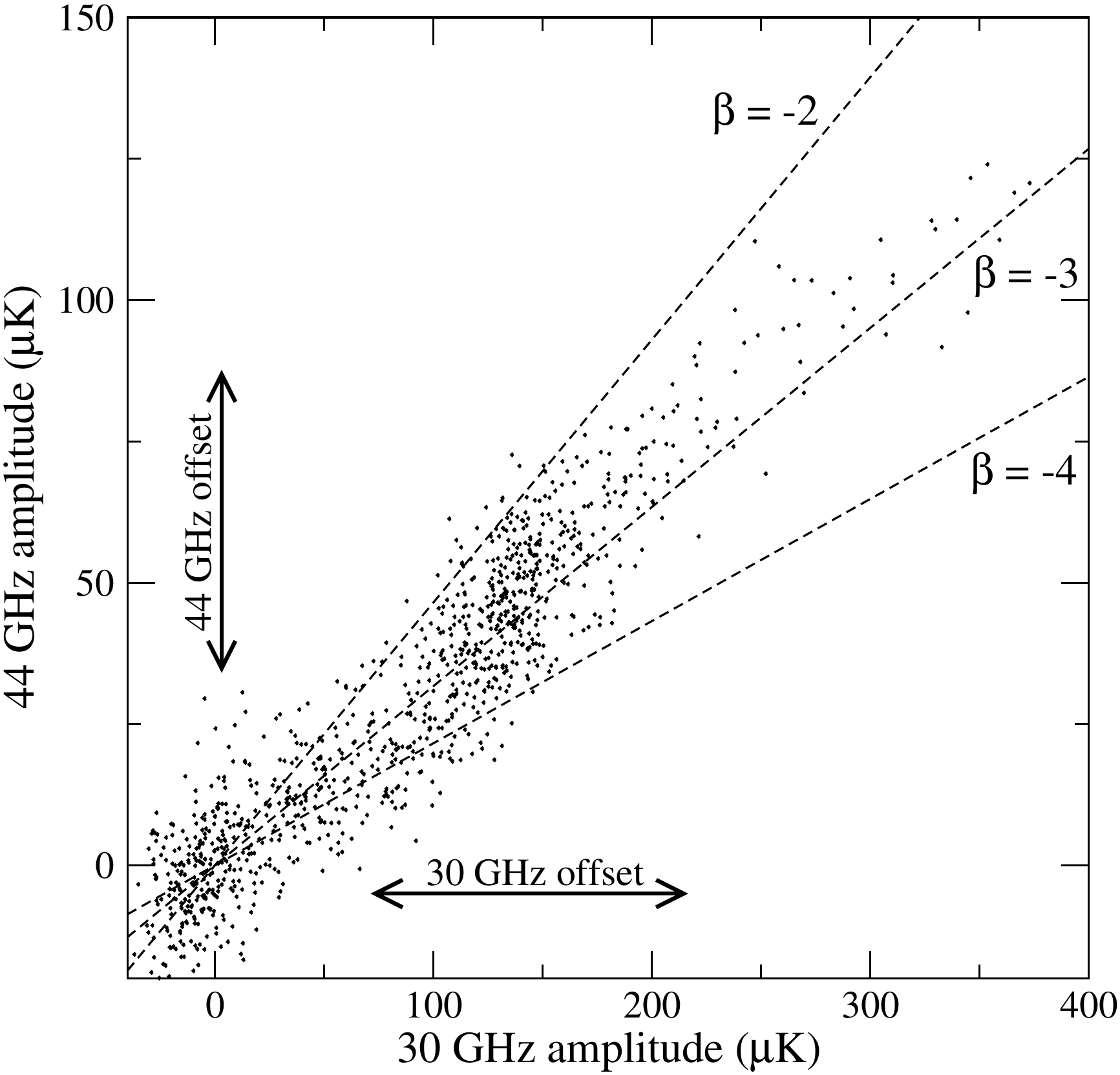,width=\linewidth,clip=}}
\end{center}
\caption{Schematic illustration of a single T--T plot evaluated for an
  ideal low-frequency \Planck\ simulation (30 and 44$\,$GHz). Each dot
  represents the observed values at two frequencies for a single
  pixel. The slope of the distribution is given by the spectral index
  of the signal component in the field, as indicated by the dashed
  lines. A constant offset in either frequency map simply translates
  the entire T--T plot either horizontally or vertically. For maps
  without spurious offsets, the best-fit straight line should pass
  through the origin; the fundamental idea of the algorithm presented in
  this paper is to ensure that this is the case for any sufficiently
  small patch of the sky.}
\label{fig:cartoon_single}
\end{figure}

In the following, we exploit the same idea to estimate both spurious
monopoles and dipoles by noting that when the true values are
correctly determined, the regression intercept has to be zero: If the
foreground signal at one frequency is exactly zero, it also has to be
exactly zero at the other frequency\footnote{Line emission processes,
  such as that arising from carbon-monoxide (CO), are clearly
  exceptions to this, and regions with significant line emission must
  be masked before applying the method presented in this paper.}. This
implies that a single T--T plot constrains both the spectral index of
the component and the relative offsets of the two maps, $m_1$ and
$m_2$, to $m_1 = am_2 + b$, where $a$ and $b$ are the slope and
intercept of the T--T plot. The main goal of the present paper is to
develop this simple idea into a complete and robust method for
determining both monopoles and dipoles from a set of multi-frequency
sky maps.

\section{Method}
\label{sec:method}

Before describing the main method, we note that robust linear
regression is in general difficult in the low signal-to-noise regime,
as both the slope and intercept are associated with large
uncertainties. We therefore adopt a two-step process in which we
first use the main T--T plot method for high signal-to-noise frequency
channels, followed by a direct template fit method for low
signal-to-noise channels. The second stage, however, is both
conceptually and implementationally straightforward and well
established in the literature; the difficult task is to set the
offsets correctly for the high signal-to-noise components, and this is
our primary concern in the following.

\subsection{Single region data model}

We start by considering a basic data model consisting of a single
signal component on the form
\begin{equation}
d_{\nu}(p) = F(\nu) s(p) + n_{\nu}(p) + m_{\nu},
\label{eq:model1}
\end{equation}
where $d_{\nu}(p)$ denotes the observed sky map value at frequency
$\nu$ and pixel $p$, $s$ represents the true sky signal (at
some reference frequency, $\nu_0$), $F(\nu)$ denotes the (normalized)
frequency spectrum of the sky signal (often called mixing matrix in
the component separation literature), $n_{\nu}(p)$ is instrumental
noise, and $m_{\nu}$ is the unknown spurious offset we want to
estimate and remove. For now, $\m_{\nu}$ is taken to be a pure
constant (monopole), and $F(\nu)$ is assumed constant over the entire
observed field.

For many radio astronomy applications, the signal spectrum can be
approximated by a power-law, $F(\nu) = (\nu/\nu_{0})^\beta$, at least
over some limited frequency range. The effective spectral index,
$\beta$, can be computed between any two frequencies, $\nu_1$ and
$\nu_2$, using the so-called ``T--T plot'' technique. Let us first
consider the noiseless case without offsets. For this case, the
spectral index is given by
\begin{equation}
d_2 = \left(\frac{\nu_2}{\nu_1}\right)^\beta d_1
\Leftrightarrow \beta = \frac{\log\left(\frac{d_2}{d_1}\right)}{\log\left(\frac{\nu_2}{\nu_1}\right)}.
\label{eq:scatter1}
\end{equation}
Defining the slope as $a \equiv (\nu_2/\nu_1)^\beta$, and including
noise and spurious offsets, this expression reads
\begin{align}
d_2 - (n_2 + m_2) &=
a\left[d_1- (n_1 +  m_1)\right] \\
d_2 &= a d_1 + (m_2 - a\,m_1) + (n_2 - a\,n_1) \\
 &\equiv a d_1 + b + n 
\label{eq:model_offset}
\end{align}
Thus, the observed signal at one frequency is related linearly to
signal at the other frequency, with a slope given uniquely by the
spectral index, and an intercept given by $b \equiv (m_2 - a\,m_1)$.

This relation is often conveniently visualized in terms of scatter (or
``T--T'') plots, as illustrated in Fig.~\ref{fig:cartoon_single}. Each
dot indicates the observed data values at the two frequencies for one
pixel, while the dashed lines indicate three models with different
values of $\beta$. A spurious offset in either frequency map
corresponds directly to a vertical or horizontal translation of the
entire scatter plot, respectively, but does not change the
slope. Therefore, the spectral index is fully insensitive to spurious
constant offsets when estimated by this T--T plot technique. However,
as shown below spurious dipoles do bias the spectral index, because
they introduce a gradient across the field, resulting in a net
additional tilt. The appropriate correction for this is discussed
below.

As long as $F(\nu) \ne 0$ for all $\nu$, it is clear that if the data
vanish in one frequency, it also have to disappear in the other. Thus,
for data free of any spurious offsets the best-fit line through the
T--T plot must pass through the origin. Correspondingly, a non-zero
regression intercept can only be due to the offset term in Eq.
\ref{eq:model_offset}, implying that $m_1$ and $m_2$ must be related
by 
\begin{equation}
m_2 - a\,m_1 = b.
\label{eq:single_offset}
\end{equation}
In other words, the true pair of offsets have to lie somewhere along
the best-fit regression line in the T--T plot. From a single scatter
plot it is impossible to determine the precise location, but this
simple linear relation nevertheless forms the core unit of our
algorithm. 

Before proceeding with the algorithm, we make a few comments
concerning the implementation of the fitting procedure for $a$ and
$b$. First, since both $d_1$ and $d_2$ are noisy quantities, the
standard method based on the normal equations does not strictly apply,
as that estimator is known to suffer from so-called \emph{attenuation
  bias}; noise in the descriptor variable, $d_1$, biases $a$ low
\citep[e.g.,][]{draper:1998}. Second, in real data sets strong
outliers occur quite frequently; a typical example in the CMB setting
is unmasked point sources. Since we will only apply this method to
high signal-to-noise data sets in the following, the second problem is
definitely more pressing for our purposes. We accordingly adopt the
non-parametric and highly robust Theil-Sen estimator in the following
\citep{theil:1950}: We estimate the slope as $\hat{a} =
\textrm{median}[(d_2(p_2)-d_2(p_1))/(d_1(p_2)-d_1(p_1))]$ evaluated
over all pixel pairs $(p_1, p_2)$, and the intercept as $\hat{b} =
\textrm{median}[d_2(p) - \hat{a}d_1(p)]$. If one wants to apply the
same method to low signal-to-noise observations, a more appropriate
estimator is the Deming estimator \citep{deming:1948}, which properly
accounts for uncertainties in both directions. We have implemented
both of these estimators in our codes, and find fully consistent
results for the cases considered in this paper.

\subsection{Absolute monopole and dipole determination by spectral variations}

Above we assumed that all pixels in the region of interest have the
same spectral index, and the entire map may therefore be analyzed
within a single T--T plot. In reality, the true spectral index varies
across the sky to some extent. For instance, the spectral index of
synchrotron emission typically ranges between, say,
$\beta_{\textrm{s}}=-2.6$ and $-3.1$, while the thermal dust
emissivity ranges between, say, $\beta_{\textrm{d}}=1.3$ and 1.8
\citep[e.g.,][and references therein]{planck_compsep}. To account for
these variations, it is therefore useful, if not necessary, to segment
the sky map into disjoint regions such that the index can be
approximated as nearly constant within each region. In the absence of
a physically motivated segmentation algorithm, a good first-order
approximation is simply to partition the sky into some coarse
grid. For applications adopting the HEALPix pixelization
\citep{gorski:2005}, such as ours, the nested ordering scheme proves
particularly useful for this purpose, as it allows fast and localized
grid coarsening. For instance, while each sky map considered here are
pixelized on a grid built up with $14\arcm$ pixels (corresponding to a
HEALPix resolution parameter of $N_{\textrm{side}}=256$), a single
constant-index region is typically defined in terms of $3\pdeg8$
pixels ($N_{\textrm{side}}=16$), each containing $16\times16$
high-resolution pixels.

Such partitioning into smaller regions is not only useful in order to
ensure nearly constant spectral indices, but it allows in fact an
absolute determination of the individual offsets, $m_1$ and $m_2$. Let
the full data set be partitioned into $N$ regions, each with a
(nearly) constant spectral index $\beta_i$, and perform an independent
linear regression procedure for each region, as outlined above. In
this case, one will obtain an independent linear constraint on $m_1$
and $m_2$, given by Eq.~\ref{eq:single_offset}, \emph{from each
  region}. These may be combined into the following (over-determined)
linear system,
\begin{equation}
\left[
\begin{array}{cc}
-a_1 & 1 \\
-a_2 & 1 \\
\vdots& \\
-a_N & 1 \\
\end{array}
\right] 
\left[
\begin{array}{c}
m_1 \\
m_2
\end{array}
\right] 
=
\left[
\begin{array}{c}
b_1 \\
b_2 \\
\vdots \\
b_N
\end{array}
\right] 
\label{eq:offset_multi}
\end{equation}
Writing this linear system in a matrix form, $\mathbf{A}\mathbf{m} =
\mathbf{b}$, it has a unique solution given by the normal equations,
$\hat{\mathbf{m}} = (\mathbf{A}^t\mathbf{A})^{-1} \mathbf{A}^t
\mathbf{b}$. Thus, any spatial variation in the spectral index breaks
the degeneracy between the offsets at the two frequencies, and at
least formally allows absolute determination of both.

\subsection{Joint monopole and dipole determination}

Partitioning the sky into sub-regions further allows us to estimate
additional degrees of freedom. Specifically, suppose that the total
offset parameter space may be spanned by some set of basis vectors,
$T_k(p)$, each with an unknown amplitude, $z_k$. The archetypal
example is the space spanned by a monopole and three dipoles, such
that the total effective offset for a given frequency map is
\begin{align}
\tilde{\mathbf{m}} &= \mathbf{T}\mathbf{z} \\
&=\left[
\begin{array}{cccc}
1 & \cos \phi_1 \, \sin \theta_1 & \sin \theta_1 \sin \phi_1 & \cos \theta_1
\\
1 & \cos \phi_2 \, \sin \theta_2 & \sin \theta_2 \sin \phi_2 & \cos \theta_2
\\
\vdots&\vdots & \vdots & \vdots\\
1 & \cos \phi_n \, \sin \theta_n & \sin \theta_n \sin \phi_n & \cos \theta_n 
\end{array}
\right]
\left[
\begin{array}{c}
z_0\\
z_1\\
z_2\\
z_3
\end{array}
\right],
\end{align}
where subscripts indicate pixel number. 
To account for these new degrees of freedom within a single region $i$,
Eq.~\ref{eq:single_offset} generalizes to
\begin{equation}
\sum_{k=0}^{3} T_k(i) m_{k,2} - a\,\sum_{k=0}^3 T_k(i) m_{k,1} = b,
\label{eq:dipole}
\end{equation}
where we approximate the net impact of the additional templates as
constant over the region, i.e., $T_k(i) = 1/N_{\textrm{pix}} \sum_{p
  \in i} T_k(p)$.  The full joint all-regions linear system,
corresponding to Eq.~\ref{eq:offset_multi}, is correspondingly
generalized into $\mathbf{A} \mathbf{x} = \mathbf{b}$, where
$\mathbf{A}$ now is an $N\times8$ matrix containing $-a_i T_k(i)$ for
region $i$ and template component $k$ in the first four columns, and
$T_k(i)$ in the four last columns. The $\mathbf{x}$ vector has 8
elements containing the template amplitudes for the first map in the
first four entries, and the template amplitudes for the second map in
the four last entries. The right-hand side, $\mathbf{b}$, is identical
to that in Eq.~\ref{eq:offset_multi}. Again, this system is
solved by the normal equations, $\hat{\mathbf{x}} =
(\mathbf{A}^t\mathbf{A})^{-1} \mathbf{A}^t \mathbf{b}$.

\begin{figure}[t]
\begin{center}
\mbox{\epsfig{figure=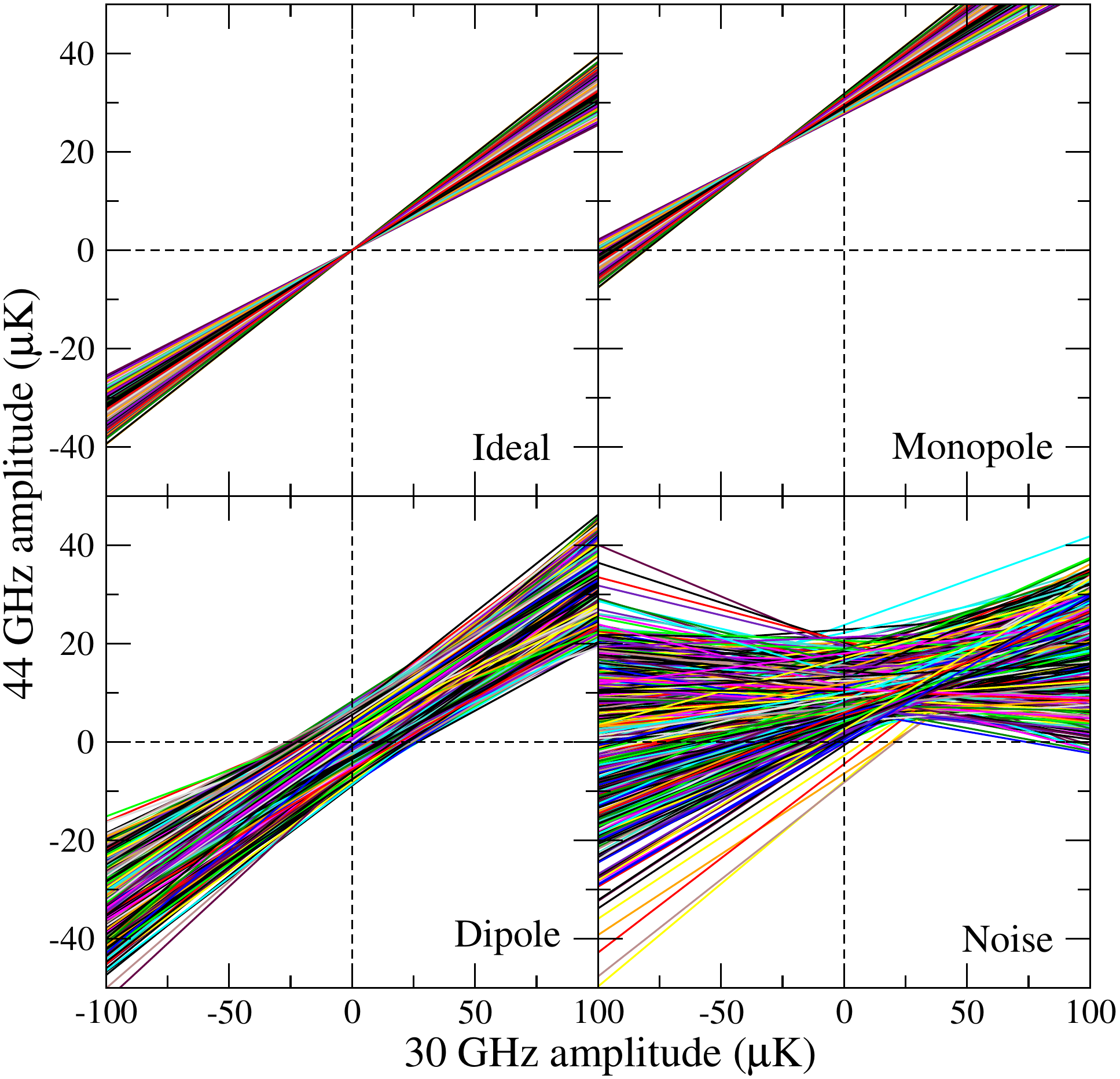,width=\linewidth,clip=}}
\end{center}
\caption{T--T summary plot of four different simulations. Each
  coloured line corresponds to the best-fit to a single HEALPix
  $N_{\textrm{side}}=8$ region, evaluated from underlying an
  $N_{\textrm{side}}=256$ synchrotron-only simulation. The top left
  panel shows the ideal case with neither spurious offsets nor
  instrumental noise; all lines converge perfectly on the origin. In
  the top right panel spurious offsets of $-30$ and +20$\,\mu\textrm{K}$
  are added to the two frequencies, resulting in a simple
  corresponding translation of the entire plot. The bottom left panel
  shows the effect of a spurious dipole; the lines no longer converge
  on a single point, as each region is effectively translated by a
  different offset. Finally, the bottom right panel illustrates the
  effect of instrumental noise. }
\label{fig:cartoon}
\end{figure}

While spurious monopoles do not change the net slope of a T--T plot,
spurious dipoles do. And the larger area the considered region covers,
the larger the effect is. We account for this effect by
iteration. That is, we 1) solve Eq.~\ref{eq:dipole} as described
above, 2) subtract the derived monopole and dipole estimates from the
raw input maps, and 3) iterate until all offset updates are smaller
than, say, 1\% of the total value. Typically three or four iterations
are needed to reach convergence.

In Fig.~\ref{fig:cartoon} four different simulations illustrate the
various cases discussed so far. Each line corresponds to the best-fit
linear fit to the T--T plot of a synchrotron-only sky evaluated at 30
and 44$\,$GHz, corresponding to the two lowest \Planck\ frequencies. Each
sky map has an angular resolution of $1\deg$ FWHM, and is
pixelized on an $N_{\textrm{side}}=256$ HEALPix grid. The spectral
index is chosen to be constant within each $N_{\textrm{side}}=8$
pixel, and drawn from a Gaussian distribution with $\beta=-3\pm0.2$,
resulting in 768 independent T--T plots.

In the top left panel, we show the ideal case with neither spurious
offsets nor instrumental noise. In this case, we see that all lines
truly converge on the origin, as they should. In the top right panel we
have added a $-30\,\mu\textrm{K}$ offset to the 30$\,$GHz channel, and a
$+20\,\mu\textrm{K}$ offset to the 44$\,$GHz channel. The entire plot is
translated accordingly, now focusing on the point $(d_1,d_2) =
(-30,20)\,\mu\textrm{K}$. Intuitively, the main goal of monopole
correction is to re-center the focal point on the origin.

Next, the bottom left panel shows the effect of adding a spurious
dipole to each frequency band. For a single region, this is almost
equivalent to a simple translation, just like a monopole; however,
each scatter plot is translated differently, depending on its position
on the sky. When considering all T--T plots simultaneously, the
overall distribution therefore appears smeared out, and possibly
offset from the correct position, depending on the relative
orientation between the two dipoles and the dominant Galactic
signal. The intuitive goal of dipole correction is to make the focus
point of this plot as sharp as possible.

Finally, the bottom right panel illustrates the effect of instrumental
noise. This simply smears out each individual line, making it harder
to assess where the lines converge to a single point. When the
instrumental noise becomes comparable to the signal, robustly
estimating the slopes and intercepts becomes very difficult, and we
choose for now not to be aggressive in this respect; for low
signal-to-noise cases, we find that simpler template fitting methods
yield more robust results.

\subsection{Preparing for real-world applications}
\label{sec:realworld}

The algorithm presented in the previous section constitutes the
central engine in our method, and is already at this stage a
self-contained and complete method for ideal data sets. However, real
data are seldom ideal, and several adjustments and extensions are
usually required before the method becomes practical. In this section
we present a list of these issues, as well as their solutions.

\subsubsection{Analyzing multi-resolution data sets}

First, most multi-frequency data sets typically have different angular
resolutions at different frequencies. In order to estimate the
spectral indices (i.e., slopes) reliably across frequencies, it is
therefore necessary to smooth all bands to a common angular
resolution. To do so, we decompose each sky map into spherical
harmonics, $d(p) = \sum_{\ell m} a_{\ell m} Y_{\ell m}(p)$. According
to the spherical convolution theorem, a convolution in pixel domain
translates into a multiplication in harmonic domain by the convolution
theorem, such that if $b_{\ell}^0$ is the Legendre transform of the
intrinsic instrumental beam, and $b_{\ell}$ is the Legendre transform
of the desired common beam, the smoothed map is given by $\hat{d}(p) =
\sum_{\ell m} (b_{\ell}/b_{\ell}^0) a_{\ell m} Y_{\ell m}(p)$.

This degradation does not change the monopole or dipoles of the
original map, and the derived low-resolution offset corrections can
therefore be applied also to the full-resolution data set. However, it
is important to note that information \emph{is} lost in this process,
in the sense that the T--T plots exhibits a smaller dynamic range
after smoothing, effectively making it harder to pinpoint the optimal
solution. One should therefore not smooth more than necessary to bring
the frequency maps to a common resolution.

\subsubsection{Joint analysis of multiple frequency bands}

Second, many recent data sets have more than two frequency channels,
whereas the T--T plot method intrinsically only involves pairs of
maps. To deal with multiple maps, we order the maps according to
frequency, and derive Eq.~\ref{eq:offset_multi} for each pair of
neighbouring frequencies. Considering the simplest case with only a
monopole degree-of-freedom for each of $k$ frequency bands, this
results in the following joint system,
\begin{equation}
\left[
\begin{array}{ccccc}
-a_1 &    1 & 0 & \ldots & 0 \\
0    & -a_2 & 1 & \ldots & 0 \\
0    &    0 & -a_3 & \ldots & 0 \\
\vdots & \vdots & \vdots & \vdots & \vdots \\
0    &    0 & 0 & \ldots & 1 
\end{array}
\right] 
\left[
\begin{array}{c}
m_1 \\
m_2 \\
\vdots \\
m_k
\end{array}
\right] 
=
\left[
\begin{array}{c}
b_1 \\
b_2 \\
\vdots \\
b_k
\end{array}
\right] 
\label{eq:offset_band}
\end{equation}
Generalization to dipole estimation and multiple regions is
straightforward, although somewhat notationally involved. Note also, of
course, that nothing prevents including frequency pairs beyond
neighboring in the fit; the only limitation is that any given
frequency pair should be well described by a single dominant signal
component, which typically sets an effective limit on the allowed
frequency range.

\subsubsection{Subtracting CMB fluctuations}

For frequencies between 30 and 143$\,$GHz, the \Planck\ and WMAP sky
maps are dominated by CMB fluctuations rather than diffuse Galactic
foregrounds. Over the cleanest regions of the sky, these fluctuations
can therefore in principle serve as the signal for evaluating the
scatter plot slopes and intercepts. However, this is non-trivial for
at least two reasons. First, the CMB variance is typically of the
order of $70\,\mu\textrm{K}$ on degree angular scales, while both the
desired offset precision and the instrumental noise are typically just
a few microkelvins. Second, since the CMB frequency spectrum follows a
blackbody, it has by definition a constant spectral index (equal to 0)
at all frequencies and positions on the sky. Offset estimation on CMB
fluctuations therefore give at most relative results, and even those
are associated with relatively large uncertainties.

The most straightforward solution, and the one adopted in the
following, is to subtract an estimate of the CMB sky
\citep[e.g.,][]{planck_compsep,bennett:2013} from all sky maps prior
to offset estimation. The accuracy of this estimate is not critical,
as the nature of CMB fluctuations are very close to Gaussian, and they
therefore mostly add random noise to the T--T plots. As long as the
CMB uncertainties are significantly smaller than the absolute
foreground amplitude of the relevant channel, which usually is the
case with current CMB experiments, the fit is stable. However, in
order to ensure that the resulting offset correction estimates are
directly applicable to the original sky maps, one should ensure that
whatever CMB estimate is subtracted is orthogonal to all basis
vectors, $T_k$, which in practice implies making sure that it does not
have any monopole or dipole components. However, an uncertainty in
this estimate will in the end translate directly into an uncertainty
with the very specific CMB frequency spectrum, and will therefore not
confuse component separation algorithms, as long as these also allow
for a monopole and dipole component in the CMB fit.

\subsubsection{Handling multiple signal components by masking}

\begin{figure}[t]
\begin{center}
\epsfig{figure=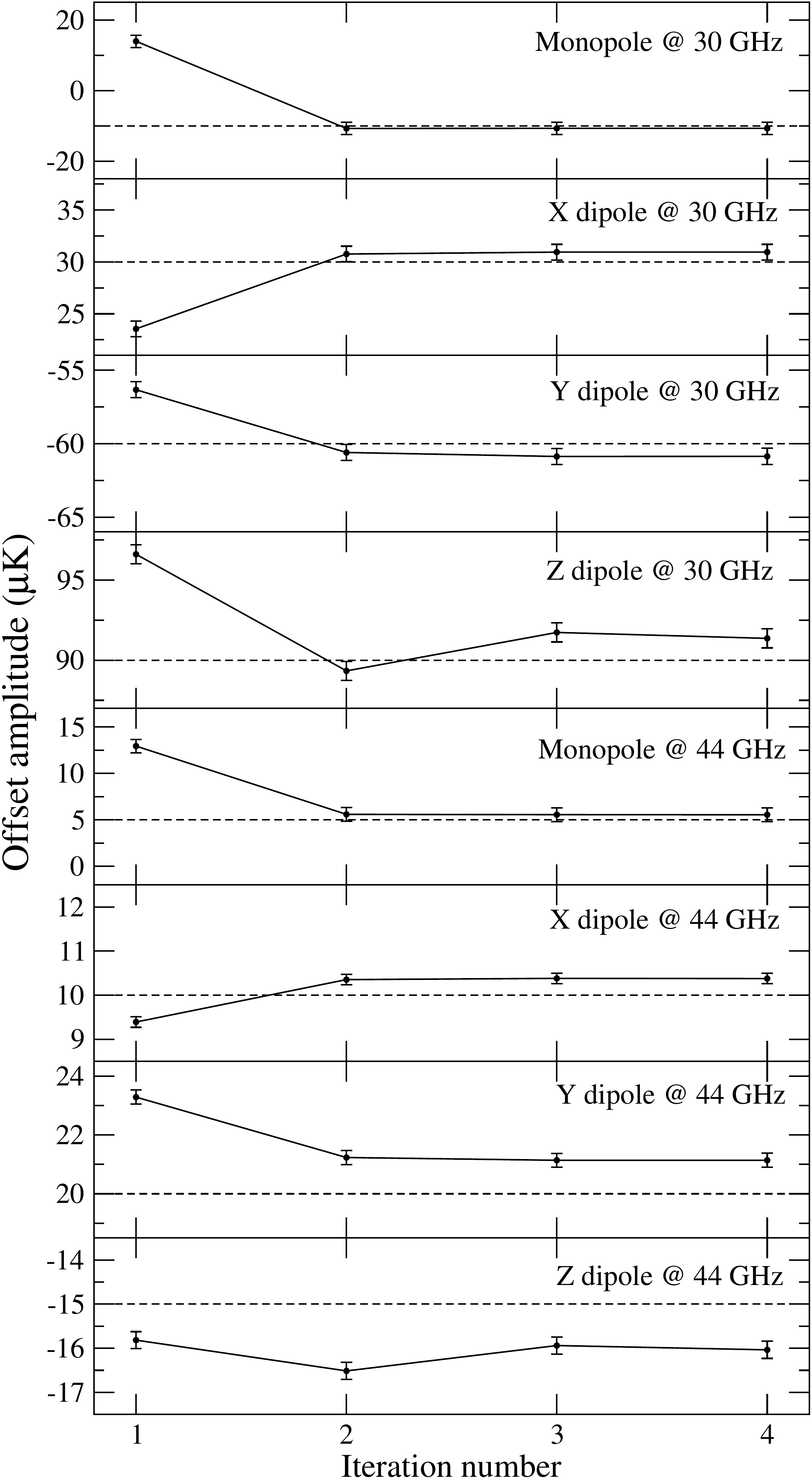,width=\linewidth,clip=}
\end{center}
\caption{Recovered offset coefficients as a function of analysis
  iteration for a two-band simulation. The true input values are shown
  as dashed horizontal lines. The values change between iterations
  because spurious dipoles bias the slope of the T--T plots; iterative
  dipole correction remove this bias. Overall, the method reproduces
  the true input values to $\lesssim1\,\mu\textrm{K}$. However, the
  adopted bootstrap uncertainties tend to underestimate the
  uncertainties in the sub-dominant channels (44$\,$GHz in this case),
  and are only intended to give an indication of the true
  uncertainties. End-to-end simulations are required for fully reliable uncertainty
  estimation.}
\label{fig:convtest}
\end{figure}

\begin{figure}[t]
\begin{center}
\epsfig{figure=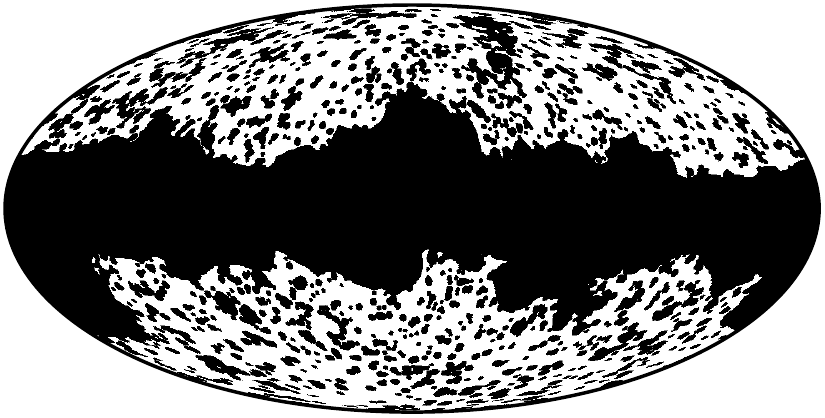,width=\linewidth,clip=}
\end{center}
\caption{Joint mask used for the analysis of the WMAP, \Planck\ and
  $100\,\mu\textrm{m}$ maps.}
\label{fig:mask}
\end{figure}

\begin{figure*}[thb]
\begin{center}
\mbox{\epsfig{figure=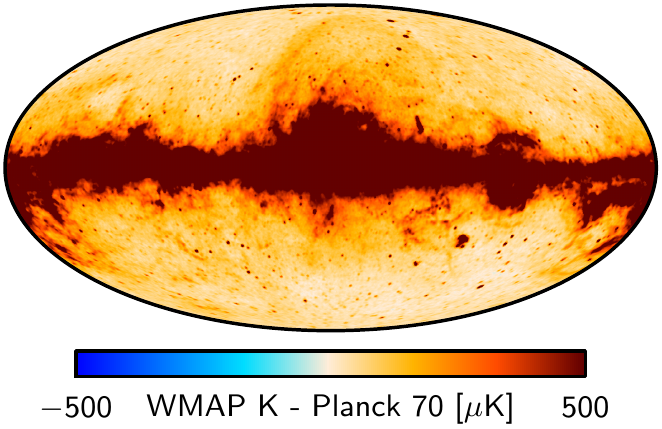,width=39mm,clip=}
      \epsfig{figure=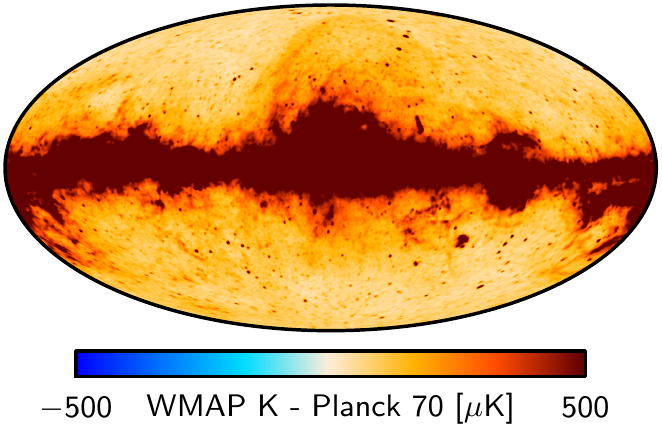,width=39mm,clip=}}
\hspace*{2mm}
\mbox{\epsfig{figure=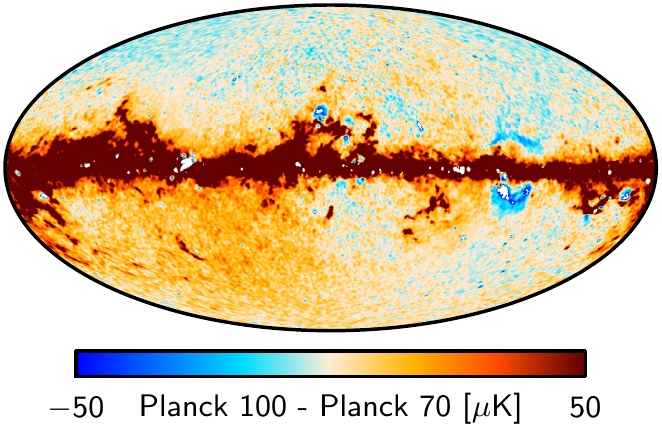,width=39mm,clip=}
      \epsfig{figure=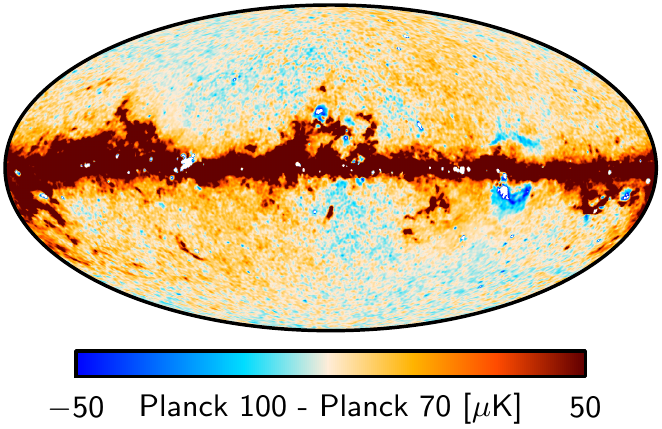,width=39mm,clip=}}
\mbox{\epsfig{figure=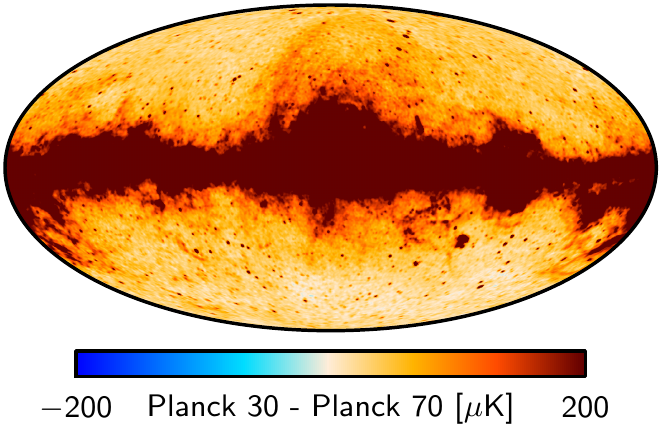,width=39mm,clip=}
      \epsfig{figure=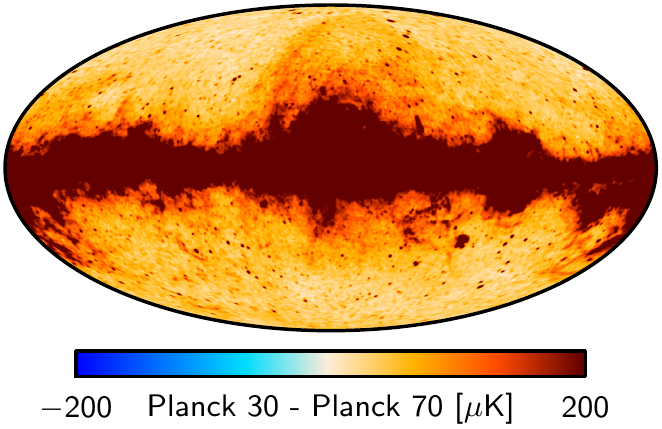,width=39mm,clip=}}
\hspace*{2mm}
\mbox{\epsfig{figure=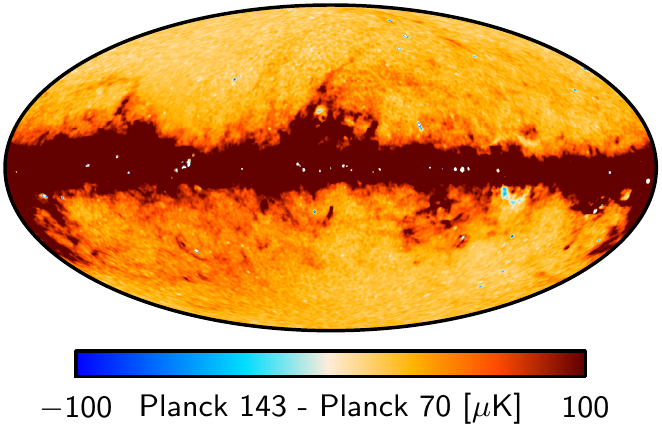,width=39mm,clip=}
      \epsfig{figure=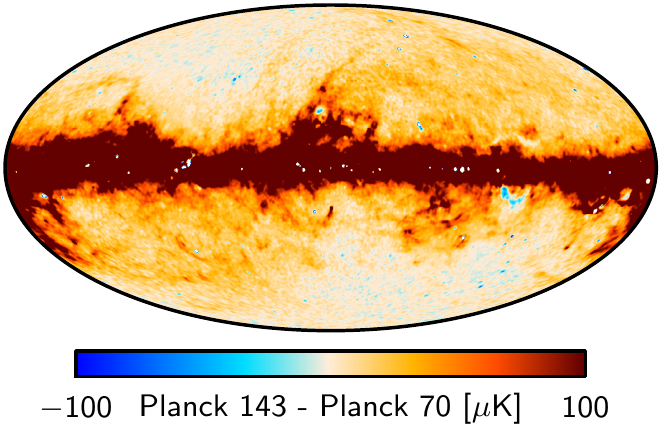,width=39mm,clip=}}
\mbox{\epsfig{figure=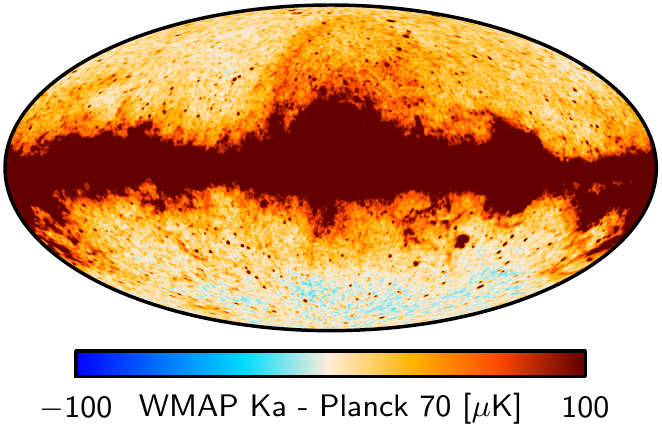,width=39mm,clip=}
      \epsfig{figure=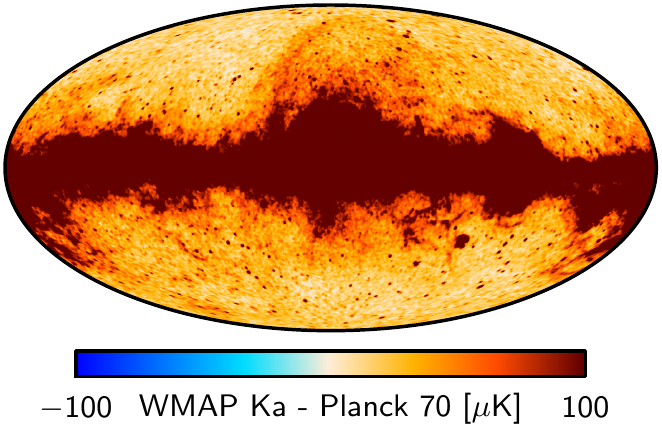,width=39mm,clip=}}
\hspace*{2mm}
\mbox{\epsfig{figure=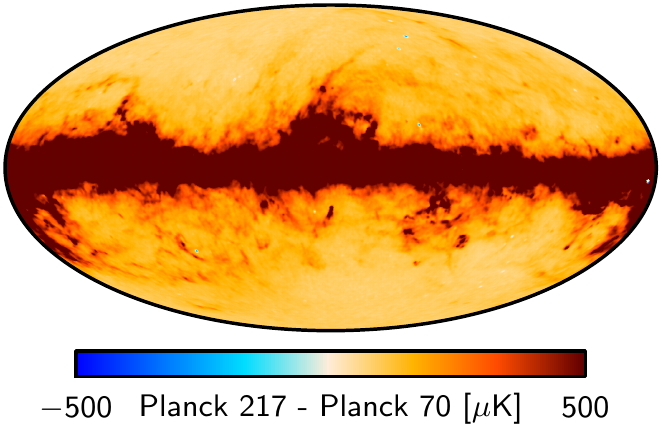,width=39mm,clip=}
      \epsfig{figure=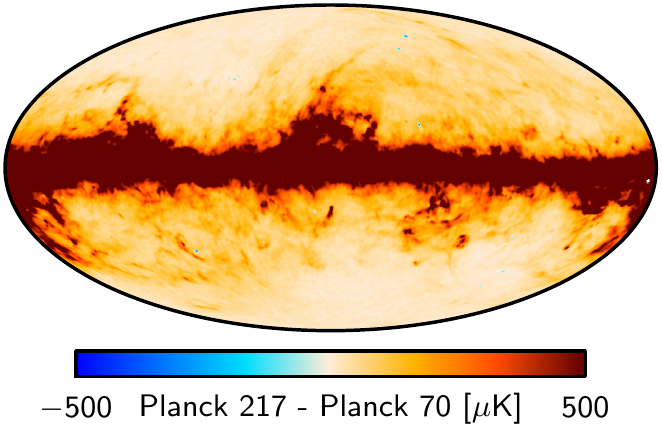,width=39mm,clip=}}
\mbox{\epsfig{figure=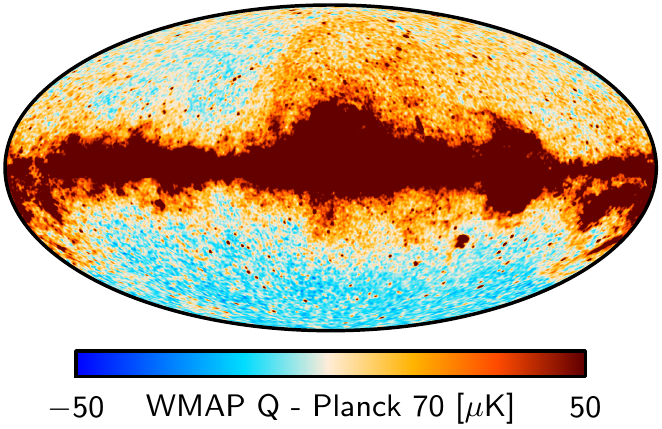,width=39mm,clip=}
      \epsfig{figure=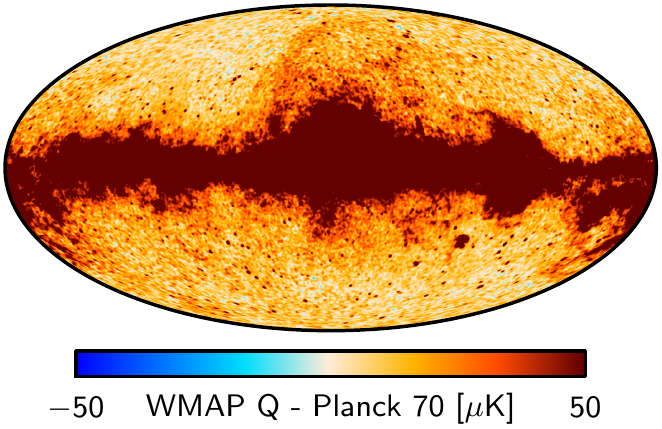,width=39mm,clip=}}
\hspace*{2mm}
\mbox{\epsfig{figure=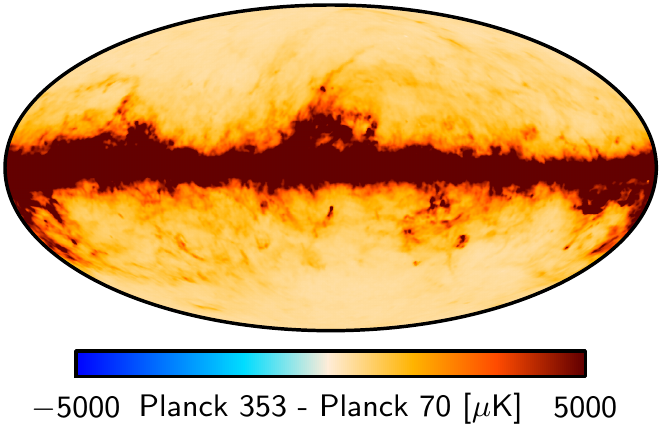,width=39mm,clip=}
      \epsfig{figure=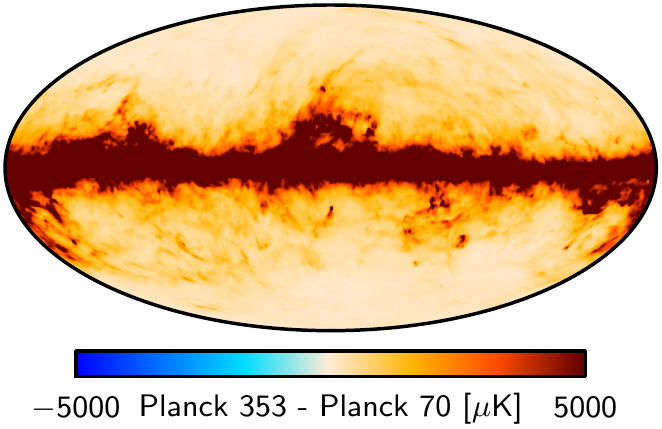,width=39mm,clip=}}
\mbox{\epsfig{figure=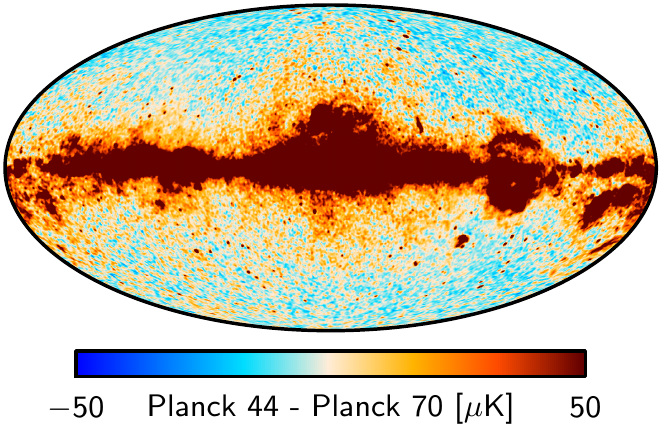,width=39mm,clip=}
      \epsfig{figure=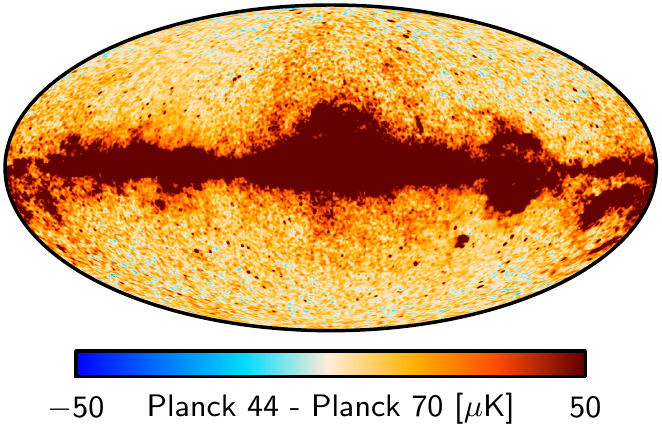,width=39mm,clip=}}
\hspace*{2mm}
\mbox{\epsfig{figure=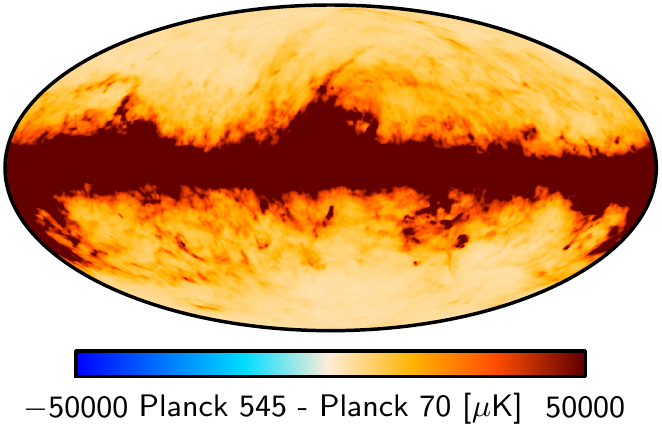,width=39mm,clip=}
      \epsfig{figure=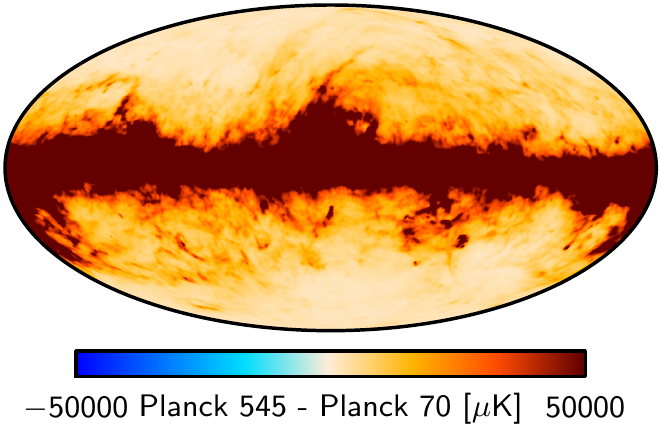,width=39mm,clip=}}
\mbox{\epsfig{figure=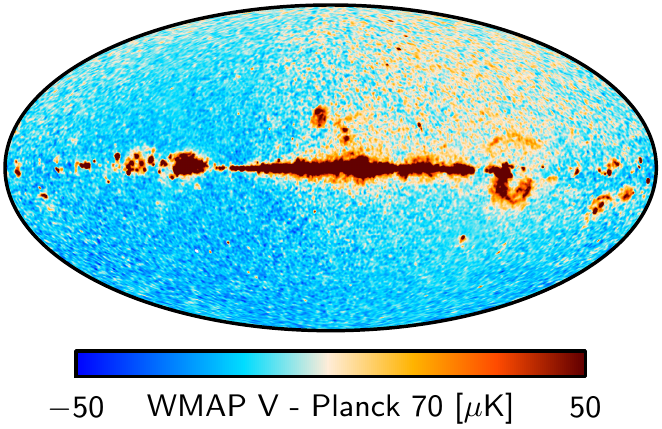,width=39mm,clip=}
      \epsfig{figure=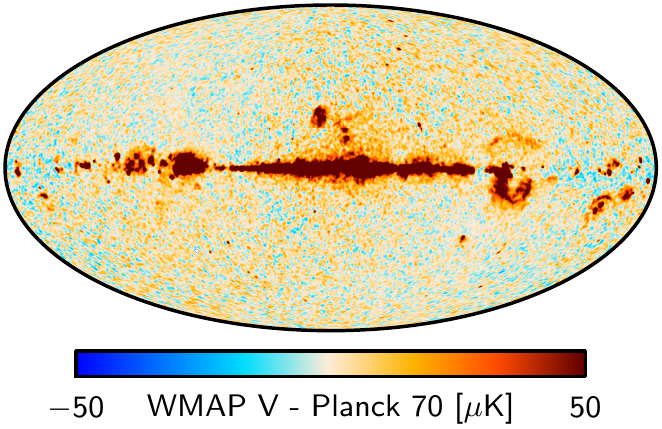,width=39mm,clip=}}
\hspace*{2mm}
\mbox{\epsfig{figure=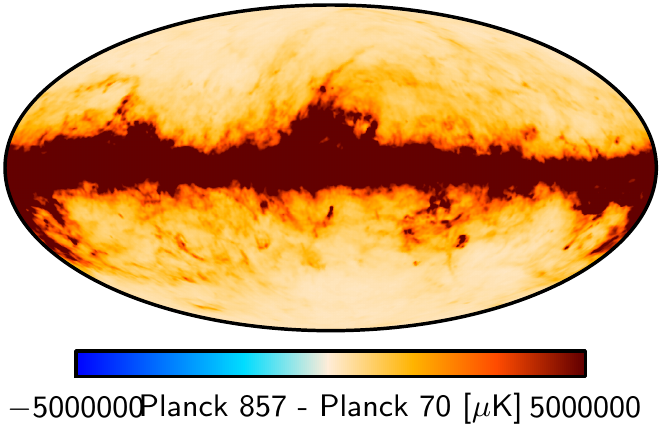,width=39mm,clip=}
      \epsfig{figure=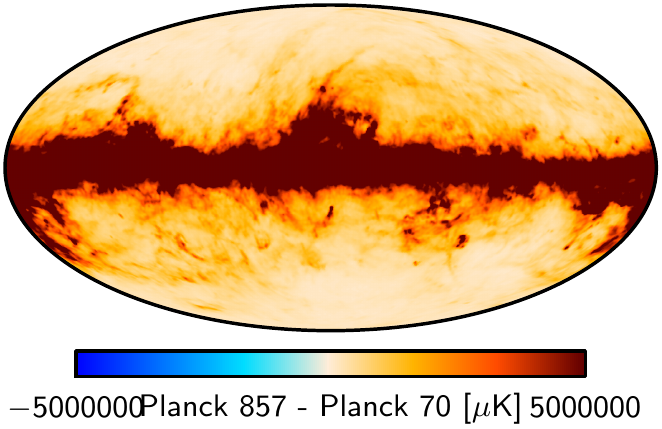,width=39mm,clip=}}
\mbox{\epsfig{figure=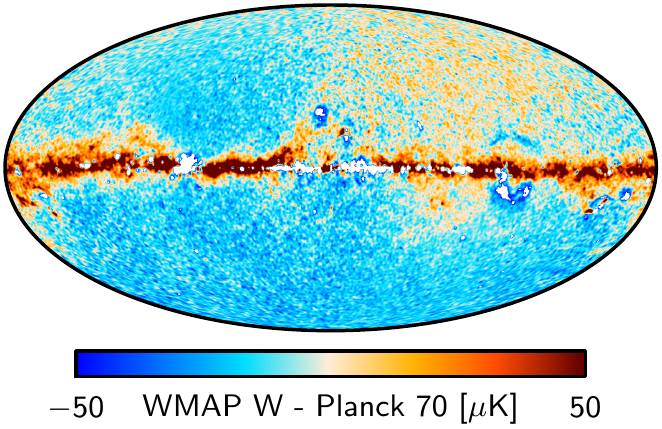,width=39mm,clip=}
      \epsfig{figure=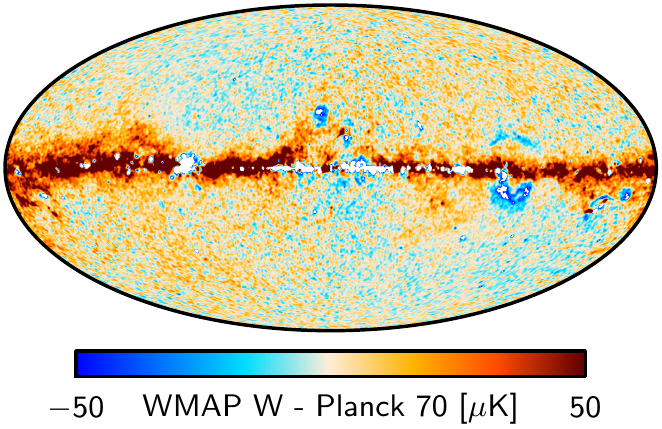,width=39mm,clip=}}
\hspace*{2mm}
\mbox{\epsfig{figure=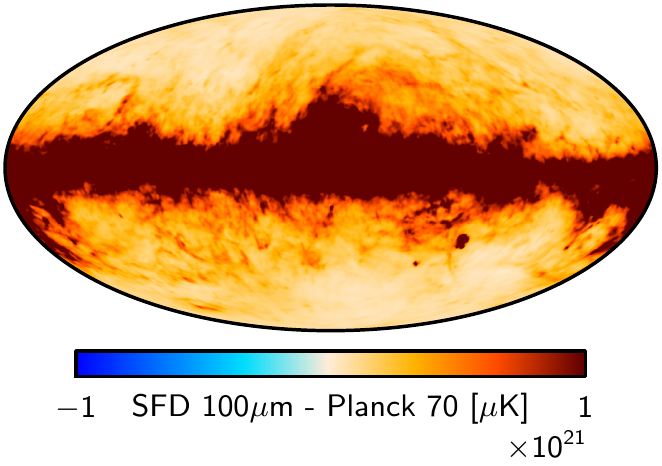,width=39mm,clip=}
      \epsfig{figure=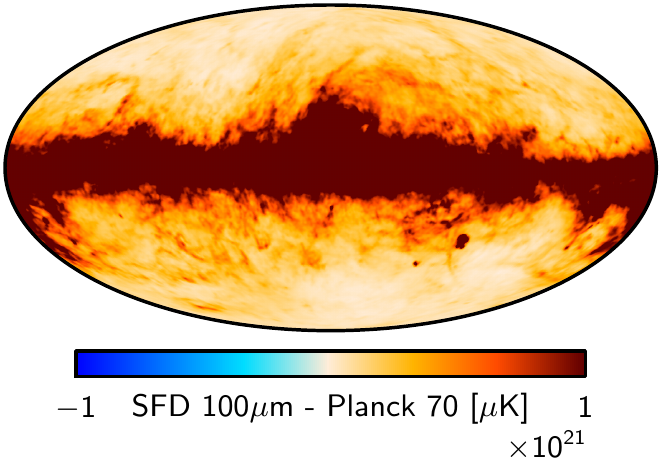,width=39mm,clip=}}
\mbox{\epsfig{figure=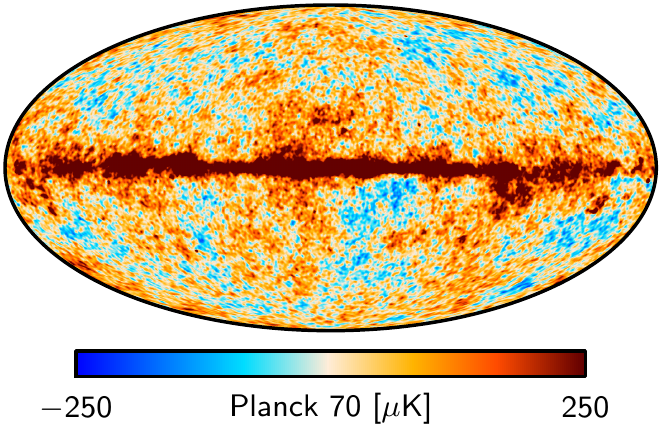,width=39mm,clip=}
      \epsfig{figure=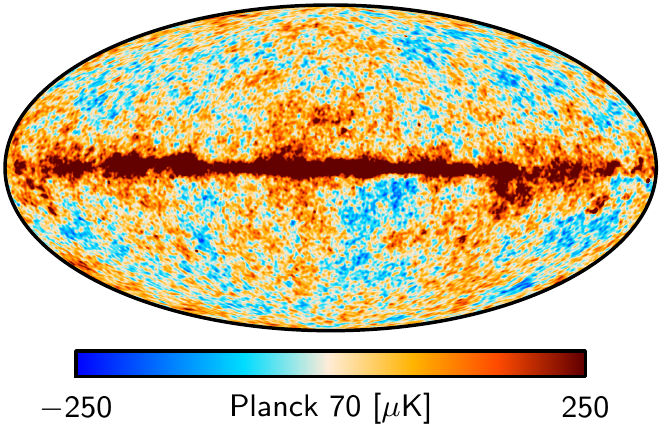,width=39mm,clip=}}
\end{center}
\caption{Difference maps between the CMB and high-frequency channels
  and the 70$\,$GHz map, before (first and third columns) and after
  (second and fourth columns) offset corrections. The bottom row shows
  the pre- and post-correction 70$\,$GHz map itself. 
}
\label{fig:diffmaps}
\end{figure*}

Related to the previous issue, the algorithm intrinsically assumes the
presence of only one dominant foreground component per region. For
data sets such as the low-frequency and synchrotron dominated
408$\,$MHz Haslam and 1420$\,$MHz Reich \& Reich maps, this is for
most parts of the sky not an issue. Neither is it for the
high-frequency \Planck\ channels above 143$\,$GHz, which are dominated
by thermal dust emission. However, for the low-foreground \Planck\ and
WMAP CMB channels between, say, 30 and 100$\,$GHz somewhat greater
care is warranted. For these frequencies, the overall signal budget is
made up by a combination of synchrotron, free-free, anomalous
microwave emission (AME; spinning dust), CO and thermal dust emission
\citep{planck_compsep}. For some frequencies, this complication can be
solved by masking, by removing spatially localized components such as
CO and free-free. For other frequencies, say, between 20 and
40$\,$GHz, where both synchrotron and AME are significant and not
spatially localized, more sophisticated component separation methods
should be used, simultaneously accounting for multiple components.

\subsubsection{Template fitting near the foreground minimum}

Near the foreground minimum around 70$\,$GHz, low-frequency
foregrounds (AME, free-free and synchrotron) and thermal
dust contribute (by definition) equally, and the scatter plot
technique is therefore intrinsically unreliable. In addition, the
foreground signal-to-noise level is low, further destabilizing the
T--T plot technique. On the other hand, the absolute foreground levels
are also correspondingly low, and even simple methods are typically
able to derive quite accurate monopole and dipole estimates. We
therefore adopt the following template fitting technique for
frequencies between 33 (WMAP Ka-band) and 100$\,$GHz: we first derive
absolute offset corrections for all high-foreground frequencies, and
apply these to the respective (CMB-subtracted) channel maps. We then
adopt the nearest neighbors on either side of the foreground minimum
(i.e., the \Planck\ 30 and 143$\,$GHz maps) as low-frequency
foreground and thermal dust templates, respectively. We then fit these
together with the monopole and three dipoles to the low-foreground
channels by solving the generalized normal equations,
\begin{equation}
\mathbf{z} = (\mathbf{T}^T\mathbf{N}^{-1}\mathbf{T})^{-1}
\mathbf{T}^T\mathbf{N}^{-1}\mathbf{d},
\end{equation}
where $\mathbf{T}$ is the $N_{\textrm{pix}}\times 6$ matrix listing
each template column-wise, and $\mathbf{N}$ is the (assumed diagonal)
pixel noise covariance matrix. A very conservative mask excluding both
point sources and residual Galactic contamination is applied in the
fit, as shown in Fig.~\ref{fig:mask}. 

We typically find that the uncertainties from this fit are on the
order of a few microkelvins. However, even such small uncertainties can in
principle be detected by detailed $\chi^2$ based analyses, for
instance as implemented in parametric foreground methods like
\commander\ \citep{eriksen:2008}. For such applications, we recommend
that one fits (or at least verifies) the offset amplitudes for these
channels within the joint analyses itself, accounting simultaneously
for foreground amplitudes and overall offsets.

\subsubsection{Stabilization by a positivity prior}

To improve the rigidity and physicality of the fit, it can be
advantageous to impose a positivity prior on the foreground
amplitudes, by requiring that the post-correction frequency map is
non-negative everywhere. We implement this as an optional feature in
our codes as follows. We locate the coldest set of pixels on the sky,
separated by at least $10^{\circ}$ on the sky. Each pixel value
defines an independent inequality constraint on the offset
coefficients given by
\begin{equation}
d(p) - \sum_{i=1}^{4} T_i(p) z_i \ge 0.
\label{eq:constraint}
\end{equation}
For cases with significant noise contribution, the inequality should be
relaxed by adding $N\sigma(p)$ to the right-hand side, where
$\sigma(p)$ is the instrumental noise RMS in pixel $p$, and $N$ is a
threshold level in units of $\sigma$; we adopt $4\sigma$ as our
positivity threshold.  Note that optimization of, say, Eq.
\ref{eq:offset_band} under these constraints is a substantially
computationally more complicated problem than solving the linear
normal equations, and must be performed using non-linear methods.

\subsubsection{Uncertainty estimation}

Finally, we make a short note on estimation of statistical
uncertainties, but emphasize that this topic is by far the most
complicated part of the entire procedure, and the method outlined here
is only intended to give a rough estimate of the uncertainties. The
fundamental problem is that for most current experiments, the monopole
and dipole coefficient uncertainties are vastly dominated by
systematic effects (foreground modelling, optical imperfections etc.),
rather than instrumental noise. Noise-based uncertainties are therefore
virtually meaningless for describing true uncertainties. For this
reason, we adopt a Monte Carlo based bootstrap method for now, aiming
to capture some of these intrinsic systematic uncertainties in a
non-parametric manner. From a data set consisting of $m$ disjoint sky
regions, we select randomly a sub-sample of $m$ regions (i.e., one
region may be included several times), and perform the full analysis
on this sub-sample in the same manner as for the original data
set. This process is repeated typically 100 times, and the resulting
variance among those 100 resamples is taken as the bootstrap
uncertainty. Further, any tunable parameter, such as whether to
perform the analysis on $N_{\textrm{side}}=4$, 8 or 16 regions, are
also drawn randomly within their allowed ranges between each resample.

The main advantage of this bootstrap approach is that it does to some
extent account for foreground modelling uncertainties and algorithmic
choices. However, it is still determined by the actually realized sky,
and can therefore only account for those variables that vary within
our actual data set; not those that are fixed for a given realization,
but in principle stochastic. The method will therefore necessarily
underestimate the true uncertainties, and we caution against interpreting
these as Gaussian 68\% confidence levels. For fully reliable
uncertainty estimation, proper end-to-end simulations (including
different foreground realizations in each simulation) are likely to be
the only truly satisfactory solution, which is outside the scope of
the current paper.

\subsection{Analysis of a simple simulation}
\label{sec:validation}

\begin{table*}[tmb]
\begingroup
\newdimen\tblskip \tblskip=5pt
\caption{Monopole and dipole estimates for two different analysis
  configurations. The top section summarizes the results for the main
  analysis of all 17 frequency maps considered in this paper, and the
  bottom section shows the results from the reduced 14 frequency data
  set employing external priors and different masks. Uncertainties are
  defined to be the maximum of the Monte Carlo-based bootstrap error
  described in Section \ref{sec:realworld} and
  $1\,\mu\textrm{K}$. Conversion between Galactic Cartesian and polar
  coordinates is given by $(l,b) = (90^{\circ}-\textrm{acos}[Z/A],
  \textrm{atan2}[Y/A,X/A])$, where $A=\sqrt{X^2+Y^2+Z^2}$ is the
  dipole amplitude. Temperatures are given in thermodynamic units.}
\label{tab:offsets}
\vskip -3mm
\scriptsize
\setbox\tablebox=\vbox{
\def\a{\rlap{$^{\rm a}$}}
\def\b{\rlap{$^{\rm b}$}}
\newdimen\digitwidth
\setbox0=\hbox{\rm 0}
\digitwidth=\wd0
\catcode`*=\active
\def*{\kern\digitwidth}
\newdimen\signwidth
\setbox0=\hbox{+}
\signwidth=\wd0
\catcode`!=\active
\def!{\kern\signwidth}
\newdimen\decimalwidth
\setbox0=\hbox{.}
\decimalwidth=\wd0
\catcode`@=\active
\def@{\kern\signwidth}
\halign{ \hbox to 1.0in{#\leaderfil}\tabskip=3em& 
    \hfil#\hfil& 
    \hfil#\hfil& 
    \hfil#\hfil& 
    \hfil#\hfil& 
    \hfil#\hfil& 
    \hfil#\hfil& 
    \hfil#\hfil \tabskip=0pt\cr
\noalign{\doubleline}
\omit  \hfill Sky map \hfill& Method& Partner(s)& Monopole& X dipole&  Y dipole& Z dipole & Unit \cr
%
%
\noalign{\vskip 4pt\hrule\vskip 4pt} 
\multispan 8 \hfil 17 band combination; tuned mask; no external priors \hfil\cr
\noalign{\vskip 4pt\hrule\vskip 4pt} 
Haslam 408\,MHz&   T--T fit&             1420\,MHz& $8.9*\pm1.3*$&$3.2*\pm1.5*$& $!0.7*\pm 1.4*!$& $-0.8*\pm 1.5*!$& K\cr
R\&R 1420\,MHz&    T--T fit&              408\,MHz& $3.28\pm0.02$&$0.07\pm0.03$& $-0.09\pm 0.03!$& $-0.08\pm 0.03!$& K\cr
WMAP K-band&       T--T fit&               30\,GHz& $*-14\pm2!**$&  $!*2\pm3*$&      $!16\pm 4!*$&     $!*7\pm 2!*$& $\mu\textrm{K}$\cr
WMAP Ka-band&      Template fit&     30 + 143\,GHz& $**!3\pm1!**$&  $!*1\pm1*$&      $!*2\pm 1!*$&     $!*7\pm 1!*$& $\mu\textrm{K}$\cr
WMAP Q-band&       Template fit&     30 + 143\,GHz& $**!1\pm1!**$&  $!*1\pm1*$&      $!*0\pm 1!*$&      $!*6\pm1!*$& $\mu\textrm{K}$\cr
WMAP V-band&       Template fit&     30 + 143\,GHz& $**!1\pm1!**$&  $!*1\pm1*$&      $*-3\pm 1!*$&     $!*5\pm 1!*$& $\mu\textrm{K}$\cr
WMAP W-band&       Template fit&     30 + 143\,GHz& $**!2\pm1!**$&  $!*2\pm1*$&      $*-4\pm 1!*$&     $!*5\pm 1!*$& $\mu\textrm{K}$\cr
\Planck\ 30\,GHz&  T--T fit&                K-band& $*!10\pm1!**$&  $*-1\pm2*$&      $!*7\pm 2!*$&     $!*8\pm 1!*$& $\mu\textrm{K}$\cr
\Planck\ 44\,GHz&  Template fit&     30 + 143\,GHz& $**!3\pm1!**$&  $!*1\pm1*$&      $!*7\pm 1!*$&     $*-4\pm 1!*$& $\mu\textrm{K}$\cr
\Planck\ 70\,GHz&  Template fit&     30 + 143\,GHz& $*!14\pm1!**$&  $!*1\pm1*$&      $!*4\pm 1!*$&     $*-3\pm 1!*$& $\mu\textrm{K}$\cr
\Planck\ 100\,GHz& Template fit&     30 + 143\,GHz& $*!15\pm1!**$&  $!*5\pm1*$&      $!*8\pm 1!*$&     $*-7\pm 1!*$& $\mu\textrm{K}$\cr
\Planck\ 143\,GHz& T--T fit&              217\,GHz& $*!34\pm1!**$&  $!*5\pm1*$&      $!*8\pm 1!*$&     $*-8\pm 1!*$& $\mu\textrm{K}$\cr
\Planck\ 217\,GHz& T--T fit&        143 + 353\,GHz& $*!84\pm1!**$&  $!*7\pm2*$&      $!*9\pm 3!*$&     $*-9\pm 2!*$& $\mu\textrm{K}$\cr
\Planck\ 353\,GHz& T--T fit&        217 + 545\,GHz& $!315\pm9!**$&  $!33\pm16$&      $!22\pm 22!$&     $-14\pm 11!$& $\mu\textrm{K}$\cr
\Planck\ 545\,GHz& T--T fit&        353 + 857\,GHz& $0.12\pm0.01$&$0.03\pm0.02$& $-0.02\pm 0.02!$&   $0.03\pm 0.01$& MJy/sr\cr
\Planck\ 857\,GHz& T--T fit&545\,GHz + 100\,$\mu$m& $0.17\pm0.03$&$0.12\pm0.05$& $-0.05\pm 0.04!$&   $0.07\pm 0.03$& MJy/sr\cr
SFD 100\,$\mu$m&   T--T fit&                857GHz& $0.24\pm0.02$&$0.25\pm0.07$& $-0.05\pm 0.01!$&   $0.02\pm 0.01$& MJy/sr\cr
\noalign{\vskip 4pt\hrule\vskip 4pt} 
\multispan 8 \hfil 14 band combination; WMAP KQ85 mask + $|b|>25^{\circ}$; fixed 353\,GHz monopole and WMAP dipoles\hfil\cr
\noalign{\vskip 4pt\hrule\vskip 4pt} 
Haslam 408\,MHz&   T--T fit&             1420\,MHz&        $7.2*$&       $1.7*$&          $!1.0*$&          $-1.2*$& $\textrm{K}$\cr
R\&R 1420\,MHz&    T--T fit&              408\,MHz&        $3.28$&       $0.04$&          $-0.10$&          $-0.12$& $\textrm{K}$\cr
WMAP K-band&       T--T fit&               30\,GHz&         $*27$&       $!0\a$&         $*!0\a*$&          $*!0\a$& $\mu\textrm{K}$\cr
WMAP Ka-band&      T--T fit&      30\,GHz + Q-band&         $*16$&       $!0\a$&         $*!0\a*$&          $*!0\a$& $\mu\textrm{K}$\cr
WMAP Q-band&       T--T fit&     Ka-band + 44\,GHz&         $*10$&       $!0\a$&         $*!0\a*$&          $*!0\a$& $\mu\textrm{K}$\cr
WMAP V-band&       T--T fit&         30 + 143\,GHz&         $**6$&       $!0\a$&         $*!0\a*$&          $*!0\a$& $\mu\textrm{K}$\cr
WMAP W-band&       T--T fit&         30 + 143\,GHz&         $**7$&       $!0\a$&         $*!0\a*$&          $*!0\a$& $\mu\textrm{K}$\cr
\Planck\ 30\,GHz&  T--T fit&                K-band&         $*29$&         $-1$&          $*-2*$&            $*!8$& $\mu\textrm{K}$\cr
\Planck\ 44\,GHz&  T--T fit&       Q-band + V-band&         $*12$&         $!1$&          $*!7*$&            $*-7$& $\mu\textrm{K}$\cr
\Planck\ 70\,GHz&  T--T fit&       V-band + W-band&         $*20$&         $!1$&          $*!7*$&            $*-7$& $\mu\textrm{K}$\cr
\Planck\ 100\,GHz& T--T fit&     W-band + 143\,GHz&         $*17$&         $!3$&          $*!8*$&            $*-7$& $\mu\textrm{K}$\cr
\Planck\ 143\,GHz& T--T fit&   100\,GHz + 217\,GHz&         $*33$&         $!2$&          $!13*$&            $-12$& $\mu\textrm{K}$\cr
\Planck\ 217\,GHz& T--T fit&        143 + 353\,GHz&         $*73$&         $!2$&          $!13*$&            $-11$& $\mu\textrm{K}$\cr
\Planck\ 353\,GHz& T--T fit&              217\,GHz&       $308\b$&         $!5$&          $!41*$&            $-33$& $\mu\textrm{K}$\cr
\noalign{\vskip 3pt\hrule\vskip 2pt}
}}
\endPlancktablewide
\endgroup
$^a$ Dipoles fixed to official WMAP values.\\
$^b$ 353\,GHz monopole fixed by HI cross-correlation \citep{planck_hfi}.\\
\end{table*}

Before applying our method to real observations, we analyze a simple
well-controlled simulation, to check that the method produces sensible
results in this case. Specifically, we once again consider a
synchrotron-plus-noise simulation, with the signal component evaluated
at the two lowest \Planck\ frequencies, 30 and 44$\,$GHz, but this
time adopting the spatial spectral index distribution derived from the
408 and 1420$\,$MHz maps in Sect.~\ref{sec:results}, and summarized in
Fig.~\ref{fig:beta}. In addition, we add spurious monopoles and
dipoles to both maps ranging between $-60$ and +90$\,\mu\textrm{K}$.

The results from this simulation are summarized in
Fig.~\ref{fig:convtest} for each offset parameter, as a function of
analysis iteration. The dashed lines show the true value, and the
uncertainties indicate the bootstrap errors described above. The
general behaviour seen in this plot is typical for all cases we have
analyzed, and therefore serves as a useful tool for building intuition
about the performance of the method. First, and most importantly, we
see that the method overall reproduces the true input values in terms
of absolute values to a precision of at worst 1--$2\,\mu\textrm{K}$.

Second, we see that the largest changes are observed between the first
and the second iteration. This is due to the already mentioned fact
that a spurious dipole introduces a bias in the effective slope (or
spectral index) of the T--T plots, and this in turn leads to a leakage
of dipole power into the monopole. However, even the first-order
dipole correction leads to a vastly improved monopole estimate, and
subsequently nearly stable results; the 30$\,$GHz monopole jumps directly
from +15$\,\mu\textrm{K}$ to the true value of $-10\,\mu\textrm{K}$ in
the second iteration. Additional iterations only change the results
with small amounts.

Third, as already stressed in Sect.~\ref{sec:realworld}, we see that
the bootstrap uncertainties do not adequately describe the true
uncertainties in the fit for all coefficients. While they do a
reasonable job for the dominant 30$\,$GHz channel, they underestimate the
uncertainties at 44$\,$GHz by up to a factor of four or five. However, it is
again important to note that the \emph{absolute} uncertainties for the same
coefficients are small. In general, the bootstrap errors tend to
produce a reasonable estimate for the dominant channels, but
underestimate the uncertainties in the sub-dominant channels. In this paper,
we will never quote uncertainties smaller than $1\,\mu\textrm{K}$ for
any component, even if the formal bootstrap uncertainty for a few
cases is as low as $0.2\,\mu\textrm{K}$.

\section{Data and analysis setup}
\label{sec:data}

We now turn our attention to a set of 17 publicly available full-sky
maps of the radio, millimeter and sub-millimeter sky, with the goal of
establishing a consistent set of offset coefficients that can be used
for multi-experiment CMB component separation analysis. We include in
the following 1) the nine \Planck\ 2013 temperature sky maps
\citep{planck_mission}, ranging between 30 and 857$\,$GHz; 2) the five
9-year WMAP temperature sky maps \citep{bennett:2013} at K- (23
GHz), Ka- (33$\,$GHz), Q- (41$\,$GHz), V- (61$\,$GHz) and W-band (94$\,$GHz); 3)
two low-frequency maps at 408$\,$MHz \citep{haslam:1982} and 1420$\,$MHz
\citep{reich:1982,reich:1986,reich:2001}, respectively; and
4) the $100\,\mu\textrm{m}$ map by \citet{schlegel:1998}. For
\Planck\ and WMAP we use the non-foreground cleaned co-added
frequency maps; for the 408$\,$MHz map, we use the version available on
LAMBDA\footnote{http://lambda.gsfc.nasa.gov} that has no filtering
applied; and the $100\,\mu\textrm{m}$ map accounts for the DIRBE and
IRAS bandpasses.  All maps are downgraded to a common resolution of
$1^{\circ}$, and repixelized at a HEALPix resolution of
$N_{\textrm{side}}=256$. No further corrections are applied to any
map.

For the positivity prior, we need a rough estimate of the instrumental
uncertainty per pixel. For WMAP and \Planck\, we base this on the
provided high-resolution noise variance maps. For the two
low-frequency and the $100\,\mu\textrm{m}$ channels we enforce a
strict positivity prior, and simply demand that there should be no
negative pixels at all.

The 408 and 1420$\,$MHz maps are analyzed together, and separately from
the other 15 maps, which are all analyzed jointly. For the
low-frequency analysis, we adopt the WMAP KQ85 mask
\citep{bennett:2013}, and for the main analysis we additionally apply
the \Planck\ GAL60 Galactic and PCCS+SZ point source mask
\citep{planck_compsep}. This joint mask is first smoothed by a $10'$
Gaussian beam, and thresholded at a value of 0.5, to remove the very
smallest source holes from the \Planck\ mask. Then it is smoothed by a
$1^{\circ}$ Gaussian beam, and thresholded at 0.99, to exclude
residual foregrounds leaking out of the mask after smoothing each
frequency map to the common resolution of $1^{\circ}$. The resulting
mask is shown in Fig.~ \ref{fig:mask}, and permits a total of 38\,\%
of the sky.

We use the main T--T scatter plot technique for the 408--1420$\,$MHz
combination, as well as for the combination of WMAP K-band and
\Planck\ 30$\,$GHz and for all frequencies above 143$\,$GHz. For the
frequencies between WMAP Ka-band and \Planck\ 100$\,$GHz we use the
template fit technique described in Sect.~\ref{sec:realworld}, and
adopt the \Planck\ 30 and 143$\,$GHz channels as foreground tracers
for low-frequency foregrounds and thermal dust, respectively. For CMB
suppression, we adopt the \Planck\ \commander\ solution from which the
monopole and dipole has been removed outside the \commander\ analysis
mask \citep{planck_compsep}. For robustness, we also performed an
analysis using the low-foreground 9-year WMAP ILC CMB map
\citep{bennett:2013}, obtaining nearly identical results.

To explore overall stability with respect to analysis choices, we
additionally analyze a subset of the above. As reported by
\citet{planck_compsep}, the spectral index of thermal dust below 353
GHz is found to be lower than the expected value of 1.3--1.7 over
extended regions of the sky. This may be explained either in terms of
systematic uncertainties in the maps, or a break in the spectral index around
353$\,$GHz, or simply a general break-down of the simple one-component
greybody model. When including the high-frequency channels, as done
above, this feature is regularized by high signal-to-noise
measurements, resulting in a far more stable fit. However, this
stability comes at a cost; if there happens to exist a second thermal
dust component, the scatter plot technique is not well defined. In
this case, it would be better for the low-frequency channels to
exclude the high-frequency channels, and thereby reduce the overall
sensitivity to this second component. The second data set considered
here therefore comprises the 14 frequency bands up to (and including)
the \Planck\ 353$\,$GHz channel. However, as noted by
\citet{planck_compsep}, this does result in a large uncertainty for
the 353$\,$GHz monopole. Therefore, we additionally fix this number to
$308\,\mu\textrm{K}$, a value determined by \citet{planck_hfi} through
cross-calibration with HI observations.

\begin{figure}[t]
\begin{center}
\epsfig{figure=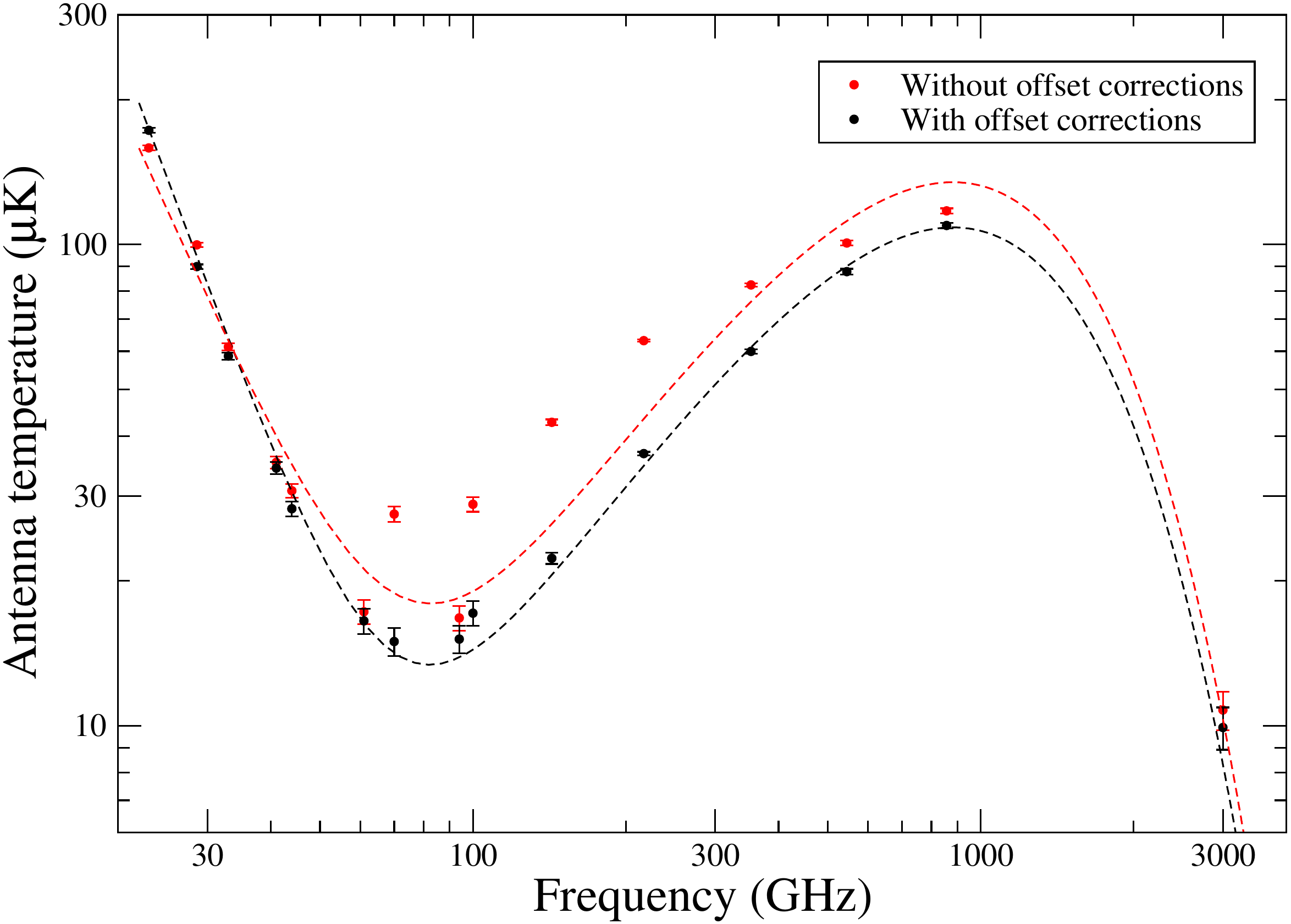,width=88mm,clip=}
\end{center}
\caption{Mean antenna temperature as a function of frequency, measured
  outside the union mask before (red points) and after (black points)
  offset corrections. The dashed lines shows the best-fit sum of a
  low-frequency power-law (with free spectral index) and a
  one-component greybody (with fixed emissivity of
  $\beta_{\textrm{d}}=1.6$ and temperature
  $T_{\textrm{K}}=18\,\textrm{K}$; Planck Collaboration XI 2013;
  Planck Collaboration XII 2013;
  Planck Collaboration Int. XVII 2014; Planck Collaboration Int. XXII
  2014) to each of the two cases.}
\label{fig:monopole}
\end{figure}

\begin{figure*}[t]
\begin{center}
\mbox{\epsfig{figure=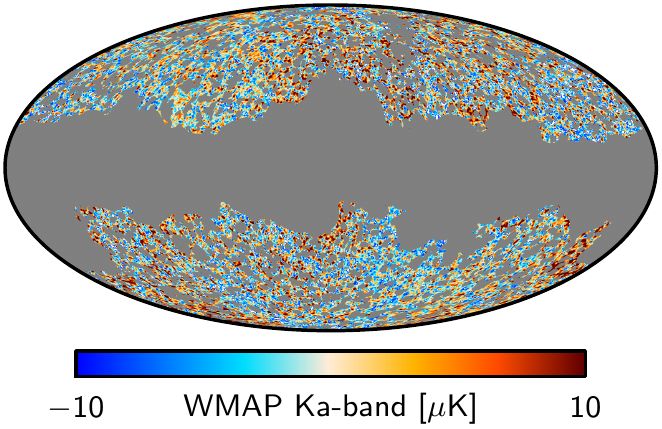,width=60mm,clip=}
      \epsfig{figure=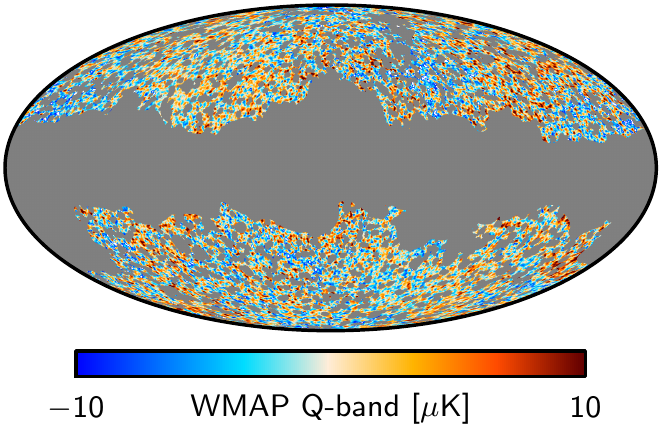,width=60mm,clip=}}
\mbox{\epsfig{figure=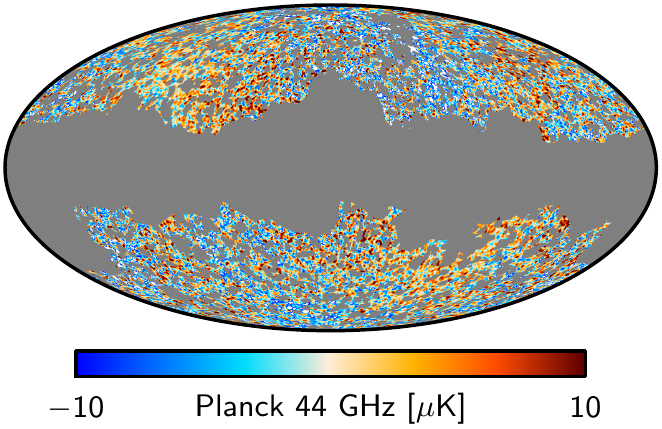,width=60mm,clip=}
      \epsfig{figure=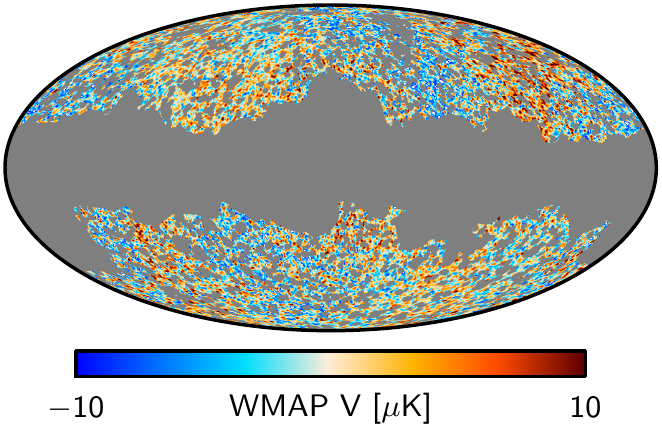,width=60mm,clip=}}
\mbox{\epsfig{figure=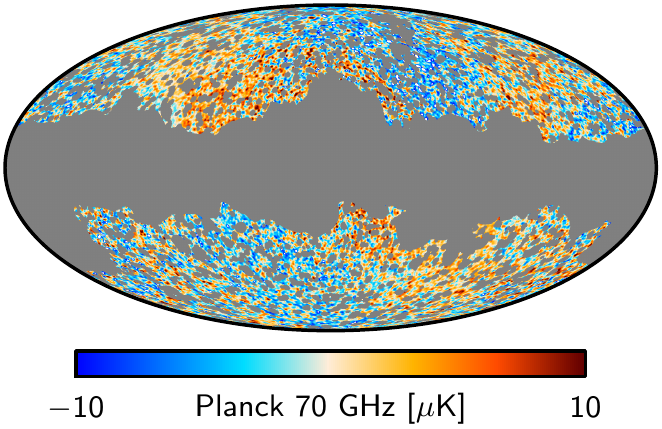,width=60mm,clip=}
      \epsfig{figure=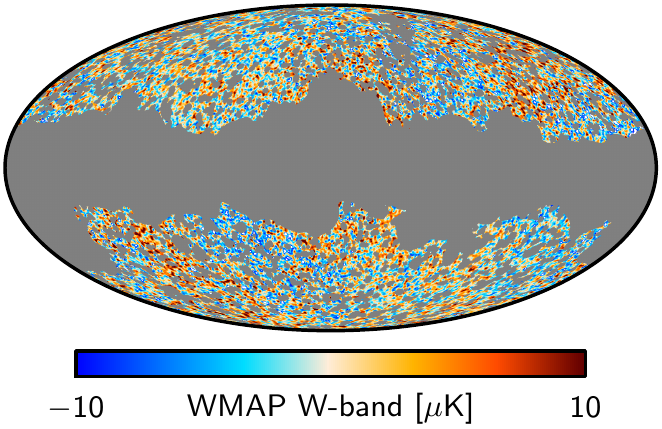,width=60mm,clip=}}
\mbox{\epsfig{figure=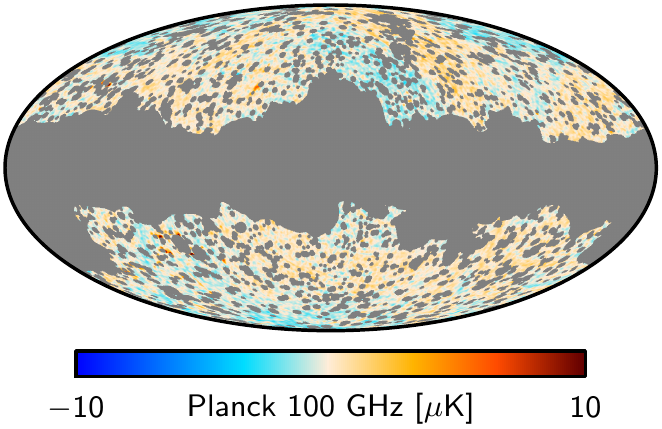,width=60mm,clip=}}
\end{center}
\caption{Map residuals after subtracting the best-fit template set
  (CMB, monopole, dipole, synchrotron/30$\,$GHz and thermal dust/143
 $\,$GHz). }
\label{fig:template_residuals}
\end{figure*}

We make three additional changes for this particular analysis. First,
we fix the WMAP dipoles to zero, recognizing that the WMAP scanning
strategy should be well suited to measure this particular
mode. Second, to probe sensitivity to sky coverage, we impose a less
conservative sky cut consisting only of the union of the WMAP KQ85
mask and a straight $|b| > 25^{\circ}$ mask, in total allowing 54\% of
the sky. In this case, we use the T--T plot technique for all
frequencies.

Analogous to fixing the WMAP dipole to zero in the consistency run,
it is of course simple to impose additional external constraints
whenever available, and these will always improve the rigidity of the
overall fit. For example, the dominant source of dipole uncertainty
for the CMB-dominated channels is the CMB dipole itself, as indeed is
demonstrated in the following. In these cases, one may therefore
impose a sharp prior on the dipole direction by including only a
single dipole template in Eq.~\ref{eq:dipole}, with a direction equal
to the CMB dipole. A second example are the high-frequency channels
above 353$\,$GHz, which are strongly signal dominated and may be assumed
to have lower relative dipole errors than the CMB channels. Imposing a
zero prior on these components may be justifiable. However, in this
paper, we fit explicitly for all components without such priors, both
to demonstrate the method and to derive minimal-assumption and
conservative estimates for all channels.

\section{Results}
\label{sec:results}

A complete summary of our main results is presented in Table
\ref{tab:offsets}, listing the monopole and dipole coefficients for
all sky maps considered in the analysis. The top section shows the
results from the main 17 frequency analysis, and the bottom section
the results from the reduced 14 frequency analysis. The uncertainties
are taken to be the maximum of the bootstrap errors discussed in
Sect.~\ref{sec:realworld} and $1\,\mu\textrm{K}$ (see Section
\ref{sec:validation}); no uncertainties are reported for the 14
frequency analysis, as the presence of external priors make these very
hard to assess.

When interpreting these results, it is important to remember that the
algorithm has been explicitly designed to leave features that are
spatially varying (i.e., Galactic structures) unchanged, while any
isotropic signal is fitted out. Thus, the fitted monopole consists of
the sum of any instrumental and data processing offsets and any
Galactic or extra-Galactic component that is spectrally uniform over
the full sky. Three typical examples are the CMB monopole of 2.73~K,
the mean value of the Cosmic Infrared Background (CIB), and the mean
generated by extra-Galactic point sources. Note, though, that the
zero-level of the Galactic foregrounds are not necessarily fitted out,
because these components have a spatially varying spectral index, and
a monopole at one frequency therefore does not correspond to a
monopole at other frequencies. Only components that are well
approximated as monopoles at \emph{all} frequencies are removed by our
fit.

In Fig.~\ref{fig:diffmaps} we show the difference between each of the
CMB and high-frequency channels with the 70$\,$GHz channel before and
after applying the monopole and dipole corrections. The 70$\,$GHz channel
is chosen as a reference because it has the lowest foreground
contamination, and each difference map should therefore ideally be
dominated by red colors. Several interesting points can be seen by eye
in this plot: First, we see significant relative dipoles in many of
the pre-correction maps. Just a few examples are WMAP Q-band,
\Planck\ LFI 44$\,$GHz, and \Planck\ HFI 100$\,$GHz. After applying our
dipole corrections, no such clear features are seen any more. Second,
we see that several of the maps are dominated by blue colors,
suggesting a relative monopole offset. Again, after applying our
monopole corrections, such features are no longer visible.

In Fig.~\ref{fig:monopole} we plot the mean temperature of the same
maps outside the union mask adopted in this analysis, adopting antenna
temperature units, both before and after offset corrections. The
dashed lines show the best-fit sum of a low-frequency power-law and a
high-frequency one-component greybody component to each of the two
data sets. While the pre-correction mean exhibits quite random
behaviour between channels, the post-correction mean follows quite well
the expected physical behaviour. The best-fit low-frequency power-law
indices are $\beta_{\textrm{s}}=-2.36$ and $-2.87$, respectively.

As described in Sect.~\ref{sec:realworld}, we adopt a straight
template-fit procedure for the low-foreground frequencies between 33
and 100$\,$GHz. A potential worry is therefore that a spatially
varying foreground spectral index can leak low-multipole power into
the monopole and dipole coefficients. To get an intuitive feeling for
the magnitude of this effect, we show in
Fig.~\ref{fig:template_residuals} the residual maps obtained by
subtracting the best-fit templates (CMB, monopole, dipole,
low-frequency foregrounds/30$\,$GHz and thermal dust/143$\,$GHz) from
each frequency map. While correlated large-scale features are indeed
seen in these figures, indicating the presence of spatial variations,
it is important to note that the color scale ranges between $-10$ and
$10\,\mu\textrm{K}$, and the magnitudes of these features are
therefore small.

\begin{figure*}[t]
\begin{center}
\mbox{\epsfig{figure=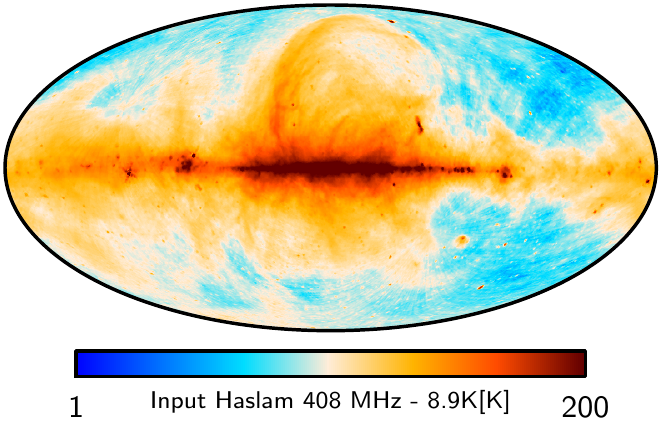,width=88mm,clip=}
      \epsfig{figure=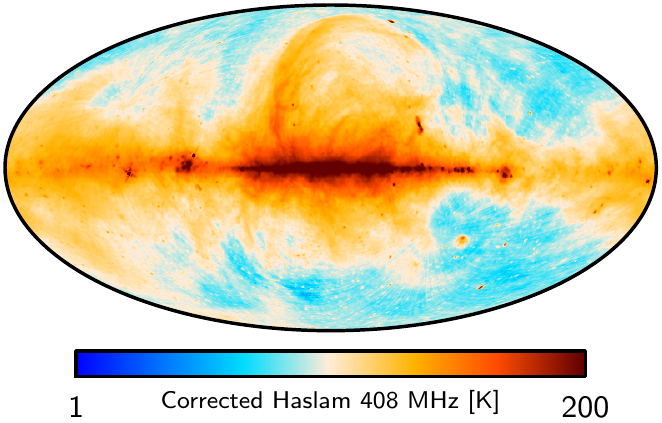,width=88mm,clip=}}
\mbox{\epsfig{figure=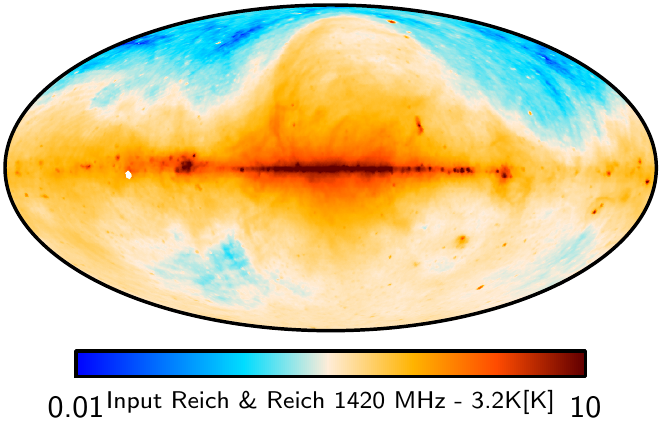,width=88mm,clip=}
      \epsfig{figure=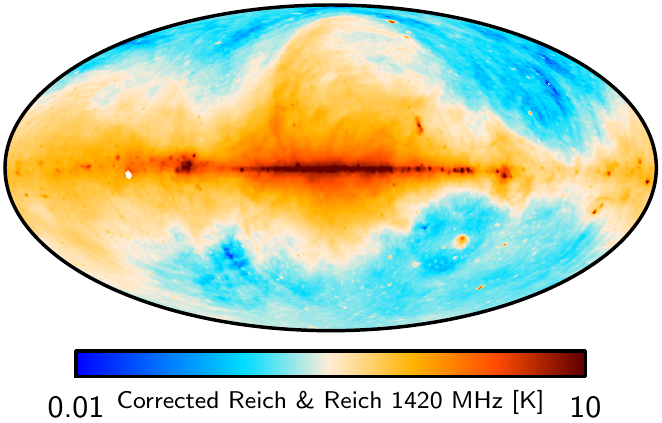,width=88mm,clip=}}
\end{center}
\caption{Comparison of the 408 (top) and 1420 (bottom) MHz
  low-frequency maps before (left) and after (right) offset
  corrections.}
\label{fig:lowfreq}
\end{figure*}

Next, in Fig.~\ref{fig:lowfreq} we compare the pre- and
post-correction 408 and 1420$\,$MHz maps. The single most striking
feature in this plot is a clear dipole extending from south to north
in the 1420$\,$MHz map. Converting the Cartesian dipole coefficients
for the 1420$\,$MHz map in Table \ref{tab:offsets} into spherical
coordinates, we find that the best-fit dipole is $0.15\pm0.03$K,
pointing towards Galactic coordinates $(l,b) =
(308^{\circ},-36^{\circ})\pm14^{\circ}$. Interestingly, this direction
is consistent with the Equatorial south pole,
$(l,b)=(303\deg,-27\deg)$, possibly suggesting that the observed
dipole might be interpreted effectively in terms of an declination
dependent offset.

After offset corrections, one can still see hints of an east-west type
dipole in both the 408 and 1420$\,$MHz maps. It is not directly
obvious whether this feature is physical or not, as the dipole X and Y
dipole coefficients for these maps are quite large (as well as
correlated), and if for instance the Y-dipole in the 408$\,$MHz is
shifted by only one standard deviation, from 0.7 to 2.1K, the visual
impression becomes far more symmetric. On the other hand, it is worth
noting that the low-frequency WMAP polarization maps shows a similar
asymmetry (see, e.g., Fig.~4 in Bennett et al. 2013), due to the
presence of strong synchrotron radiation near the Fan region.

\begin{figure}[t]
\begin{center}
\epsfig{figure=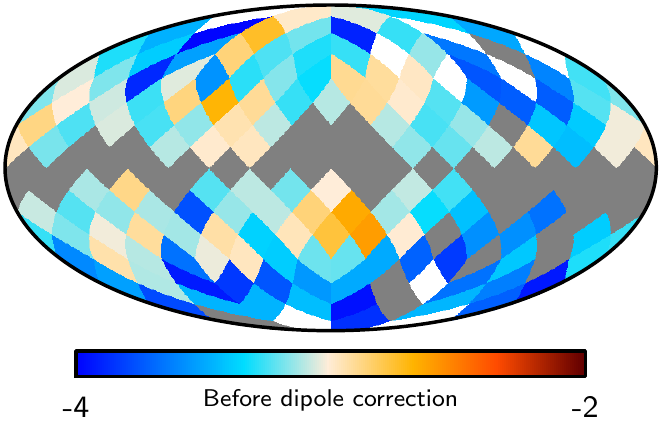,width=88mm,clip=}
\epsfig{figure=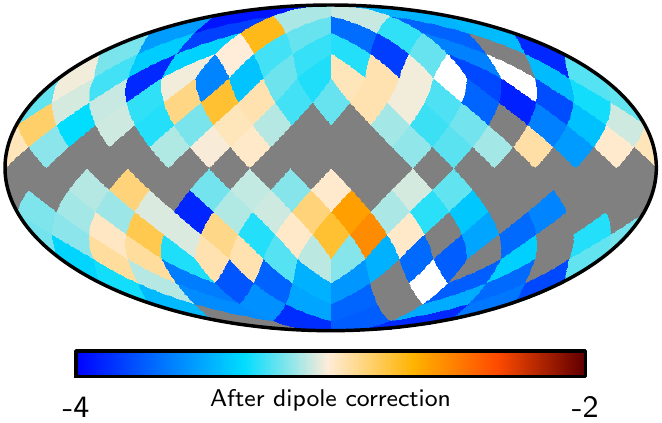,width=88mm,clip=}
\epsfig{figure=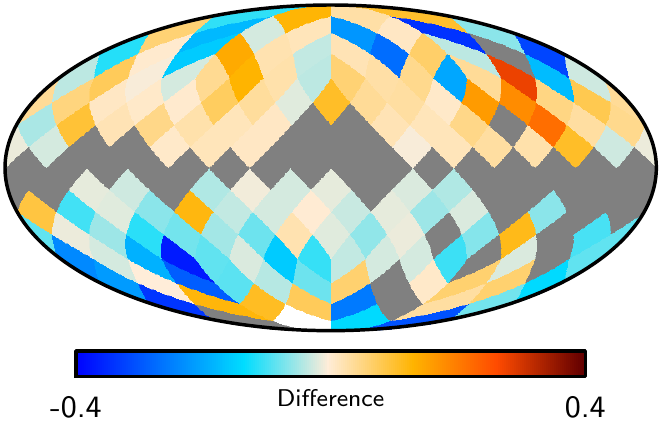,width=88mm,clip=}
\end{center}
\caption{Estimates of the spectral index derived from the
  combination of the 408 and 1420$\,$MHz maps using T--T plots, both
  before (top) and after (middle) applying the offset corrections. The
  bottom panel shows the difference. These differences show directly
  the magnitude of the bias in T--T plot based spectral indices due to
  spurious dipoles. }
\label{fig:beta}
\end{figure}

In Fig.~\ref{fig:beta} we compare the spectral index between the 408
and 1420$\,$MHz maps before and after offset corrections, as estimated
from the T--T plot distributions. Note that since monopoles do not
change the index at all, these differences are all due to the dipole
correction. Thus, the typical dipole-induced bias on the mean spectral
index as computed over $N_{\textrm{side}}=4$ regions is $\pm0.2$ for
these maps.

Finally, we consider the stability of the derived results by comparing
the results from the 17 and 14 band analyses, as presented in the top
and bottom sections of Table \ref{tab:offsets}. First, regarding the
low-frequency 408 and 1420$\,$MHz channels, we see that almost all
deviations are smaller than $1\sigma$ when including 16\% more sky
area, with the biggest difference is seen for the 408$\,$MHz monopole at
$1.3\sigma$. 

On the high frequency side, we see that the 353$\,$GHz monopole of
$315\pm9\,\mu\textrm{K}$ derived within the 17 band solution is in
excellent agreement with HI cross-correlation result of
$308\,\mu\textrm{K}$ \citep{planck_hfi}. The 353$\,$GHz dipoles shows
larger variations at the $2\sigma$ level. Imposing an HI prior also on
this component should prove useful for the 14-band analysis.

For the CMB-dominated frequencies, we see that the dipoles are overall
in good agreement in terms of absolute numbers, despite the fact that
the WMAP dipoles are forced to zero. This robustness gives added
credibility to the derived \Planck\ dipoles. On the other hand, it is
also clear that the reported uncertainties are too small, as already
seen in simulations.

Finally, we note relatively large differences in the monopole values
for frequencies between 23 (K-band) and 94$\,$GHz (W-band), both for WMAP
and \Planck. This may be linked to the relatively large K-band dipole
of $16\pm4\,\mu\textrm{K}$ in the 17-band analysis; when forcing this
value to be zero, the monopole at the same frequency increases from $-14$
to $27\,\mu\textrm{K}$, and this must in turn be accommodated by
increased monopoles at higher frequencies.

\section{Summary and conclusions}
\label{sec:discussion}

One of the most difficult tasks in physical CMB component separation
is the determination of absolute offsets, i.e., spurious monopoles and
dipoles. Unless accounted for, such offsets may bias the estimation of
spectral parameters significantly, and this can in turn lead to errors
in the actual CMB map. Ideally, the optimal approach would be to fit
the offsets and foreground model jointly, as for instance implemented
by Gibbs sampling \citep{eriksen:2008}; however, if the data model
contains a large number of spectral indices, say, one per square
degree pixel, the system is often not sufficiently rigid to uniquely
determine the optimal solution. The offsets are nearly degenerate with
the effective zero-level of each foreground component, and the only
feature that breaks this degeneracy is spatial variation in the
spectral parameters.

In this paper, we have presented an alternative method for estimating
spurious monopoles and dipoles in multi-frequency data sets. This
method builds on a well-established methodology from the radio
astronomy literature called T--T plots. The main advantages of this
method over the Bayesian approach are that 1) it makes minimal
assumptions about the nature of the signal components, 2) it is
computationally cheap, and 3) it is trivial to tune the number of
regions to the point that the degeneracy between the spectral indices
and the offsets are broken. The latter can of course also be
implemented within the Bayesian framework, at which point we expect
the two methods to perform similarly. The main disadvantage of the
current method is a relatively large systematic uncertainty when no
signal component dominates, which for CMB purposes typically happens
near the foreground minimum at 70$\,$GHz. In the present paper, we have
adopted a straight template-fit approach for these frequencies, but
note that a true multi-component fit is certainly preferable. The most
likely application for the current method is therefore to set the
offsets at the foreground-dominated frequencies, which will then serve
as an ``anchor'' for the Bayesian fit, effectively breaking the
degeneracies between the offsets and the spectral parameters.

The main products presented in the paper are two different, but each
internally consistent, sets of monopole and dipole
coefficients. Overall, the two sets agree well with each other, except
for a single common monopole extending from 23 to 94$\,$GHz. This is
largely due to the significant systematic uncertainty associated with
the WMAP K-band and \Planck\ 30$\,$GHz offsets. We recommend that
methods employing our offset values for subsequent analyses consider
both sets for systematic uncertainty assessment.

An early version of the method presented in this paper was already
adopted by the 2013 \Planck\ release to determine the zero-levels for
the physical component separation results, but only applied to the
\Planck\ frequencies between 30 and 353$\,$GHz \citep{planck_compsep}. In
this paper we have extended the total data set to also include the
WMAP frequencies, the \Planck\ 545 and 857$\,$GHz frequencies, and the
$100\,\mu\textrm{m}$ SFD map. Overall, the results presented here are
in good agreement with the earlier results, with a largest relative
monopole difference of $2\,\mu\textrm{K}$ for all channels between 30
and 217$\,$GHz. The only significant outlier is the 353$\,$GHz channel, for
which we derive a value of $315\pm9\,\mu\textrm{K}$, whereas
\citet{planck_compsep} obtained a value of $414\,\mu\textrm{K}$. A
third and fully independent estimate of this number was provided by
\citet{planck_hfi} based on cross-correlation with HI observations,
who reported a value of $308\pm23\,\mu\textrm{K}$, in excellent
agreement with our result. Similarly, at 857$\,$GHz \citet{planck_hfi}
reports a value of $0.147\pm0.0147\,\textrm{MJy/sr}$ from HI
measurements, which is to be compared with our fully internal estimate
of $0.17\pm0.03\,\textrm{mK}$; for the 545$\,$GHz the corresponding
numbers are $0.107\pm0.017$ to be compared with our value of
$0.12\pm0.01$. Other channels are also in good agreement.  However,
for the SFD 100$\,\mu$m map, we note that the X-dipole has both a
large value and a large uncertainty, suggesting a less constrained fit
overall. This is not unexpected, as this particular channel is only
coupled to the \Planck\ 857$\,$GHz map through a long frequency
extrapolation. The offset values for this channel are clearly less
robust than for the HFI channels, and its role is primarily to
stabilize the 857$\,$GHz results, rather than derive independent and
robust offsets for the SFD map itself.

On the low-frequency side, the 408$\,$MHz and 1420$\,$MHz monopoles have
already been subject of considerable discussion in the
literature. \citet{haslam:1982} estimated that the zero-level
uncertainty of the 408$\,$MHz survey was $\pm3\,\textrm{K}$, with an
additional multiplicative calibration of 10\%. The corresponding data
for the 1420$\,$MHz survey are $\pm0.5\,\textrm{K}$ and 5\%
\citep{reich:1988}. \citet{lawson:1987} used a comparison with a 404
MHz map to determine a uniform background for an assumed correct
zero-level of $5.9\,\textrm{K}$, which consists of the CMB monopole
and an isotropic extra-galactic contribution from source counts. They
also quote a zero-level correction of $-0.13\,\textrm{K}$ for the 1420
MHz survey, where the uniform background is $2.8\,\textrm{K}$. From a
TT-plot analysis \citet{reich:1988} found a $-2.1\,\textrm{K}$
zero-level offset for the 408$\,$MHz survey, when assuming that the
zero-level at 1420$\,$MHz is correct. Later, \citet{reich:2004} used an
improved source-count correction, which results in a uniform
background at 408$\,$MHz of $5.4\pm0.6\,\textrm{K}$. The zero-level
correction is $-2.7\,\textrm{K}$ then. \citet{tartari:2008} used
absolute brightness measurements at 600$\,$MHz and 820$\,$MHz and derived
zero-level offsets of $-3.9\pm0.6\,\textrm{K}$ at 408$\,$MHz and
$-0.12\pm0.14\,\textrm{K}$ at 1420$\,$MHz. All together these studies
show that the zero-level of the 408$\,$MHz survey is too low and requires
corrections between $+2.1\,\textrm{K}$ and $+3.9\,\textrm{K}$, while
the 1420$\,$MHz survey requires corrections up to $+0.13\,\textrm{K}$.
All these studies assume that the remaining extended emission in the
survey maps is of Galactic origin. Spectral index maps
\citep[e.g.,]{reich:1988,lawson:1987} do not contradict this
assumption.  However, the minima in the survey maps at 408$\,$MHz and
1420$\,$MHz, with the zero-level, isotropic source-count and CMB
corrections by the \citet{reich:2004} values, are about
$9.5\,\textrm{K}$ and $0.4\,\textrm{K}$ at 2$\degr$ angular
resolution.  This means that there is room for a Galactic contribution
to a monopole and higher order components, but also for a larger
isotropic extra-galactic component than calculated from source counts,
as discussed in \citet{sun:2010}.  Without any correction the survey
minima are $12.2\,\textrm{K}$ and $3.2\,\textrm{K}$ at 2$\degr$
resolution, respectively, which constrain the monopole components when
derived directly from the survey data.  Recently, \citet{bennett:2013}
used the same co-secant method as they applied to the WMAP CMB
frequencies to derive a background value of $7.4\,\textrm{K}$ at 408
MHz.  \citet{fornengo:2014} used six surveys to model the Galactic
synchrotron and thermal emission and found an isotropic background
value, without CMB, of $11.8\pm1.1\,\textrm{K}$ at 408$\,$MHz and
$0.58\pm0.025\,\textrm{K}$ at 1420$\,$MHz. They conclude that their
results agree with the extra-galactic component from ARCADE 2
\citep{fixsen:2011}. \citet{subramanian:2013} also modelled the
Galactic disk and halo emission and showed that the simple co-secant
method used to fit Galactic emission is the reason for the ARCADE 2
excess and that there is no need for an isotropic extra-galactic
component beyond that from source counts.  The currently deepest
source count data by \citet{vernstrom:2014}, which also include
diffuse extra-galactic emission at 1.75$\,$GHz, give a strong indication
that the excessive temperature found by ARCADE 2 is not of
extra-galactic origin. Our current monopole values of
$8.9\pm1.3\,\textrm{K}$ for 408$\,$MHz and $3.28\pm0.02\,\textrm{K}$ at
1420$\,$MHz are in the possible range allowed by the survey data.

While the method presented in this paper is very general, and can deal
with spurious monopole and dipoles of any origin, it is worth noting
that by far the dominant source of dipole uncertainty in current CMB
maps comes from estimating the solar CMB dipole. For instance, even
small calibration uncertainties can lead to a significant uncertainties in the
recovered dipole. However, this uncertainty is perfectly correlated between
channels within a given experiment, and it is therefore possible to
impose the prior that the corrections should be identical across
frequencies. 

Finally, we conclude with two important caveats. First and foremost,
it is important to realize that while the method presented here is
extremely efficient at establishing \emph{relative} offsets between
channels (which by far is the most important problem for most
component separation algorithms), it requires both high
signal-to-noise observations and significant spatial spectral
variations across the sky to determine \emph{absolute} offsets. With
the data sets studied in this paper, it appears that these criteria
hold for both the low-frequency 408 and 1420$\,$MHz maps and the
high-frequency channels above 100$\,$GHz, but not for the intermediate
CMB channels between 23 and 94$\,$GHz. Again, as shown in Section
\ref{sec:data} the systematic uncertainty on the WMAP K-band offsets
is large. More conservatively, we caution against adopting any of the
derived offsets for frequencies between 23 and 44$\,$GHz without further
cross-checks because of the presence of multiple significant
foreground components (i.e., synchrotron, free-free and AME). Instead a full
parametric fit is should be used for these channels. Second, we
emphasize that accurate uncertainty estimation is difficult, because the
dominant sources of uncertainties are generally systematic in
nature. In the present paper, we have adopted an internal bootstrap
approach, which is able to capture some, but not all, of these
uncertainties. For more realistic systematic uncertainty assessment, we
recommend propagating both offset sets provided in this paper through
subsequent analyses.

\begin{acknowledgements}
We thank Greg Dobler for useful discussions.  The development of
Planck has been supported by: ESA; CNES and CNRS/INSU-IN2P3-INP
(France); ASI, CNR, and INAF (Italy); NASA and DoE (USA); STFC and
UKSA (UK); CSIC, MICINN, JA and RES (Spain); Tekes, AoF and CSC
(Finland); DLR and MPG (Germany); CSA (Canada); DTU Space (Denmark);
SER/SSO (Switzerland); RCN (Norway); SFI (Ireland); FCT/MCTES
(Portugal); PRACE (EU). A description of the Planck Collaboration and
a list of its members, including the technical or scientific
activities in which they have been involved, can be found at
\url{http://www.sciops.esa.int/index.php?project=planck&page=Planck_Collaboration}. 
Part of the research was carried out at the Jet Propulsion Laboratory,
California Institute of Technology, under a contract with NASA. This
project was supported by the ERC Starting Grant StG2010-257080.  Some
of the results in this paper have been derived using the
\healpix\ package.
\end{acknowledgements}

\bibliographystyle{aa}

\raggedright
\end{document}